\documentclass[twocolumn,usenatbib,amsmath,amssymb,amsbsy]{mn2e}

\usepackage[T1]{fontenc}
\usepackage{aecompl}

\usepackage{natbib}
\usepackage{amsbsy}
\usepackage{amssymb}
\usepackage{amsmath}

\usepackage{graphicx}

\usepackage{color}

\begin{document}

\title[Parameters for gravitational wave searches ] {Parameter choices and ranges for continuous gravitational wave searches for steadily spinning neutron  stars}

\author[Jones]{D. I. Jones$^1$ \\
   $^1$ Mathematical Sciences and STAG Research Centre, University of Southampton, Southampton SO17 1BJ, UK}

\maketitle

\begin{abstract}
We consider the issue of selecting parameters and their associated ranges for carrying out searches for continuous gravitational waves from steadily rotating neutron stars.  We consider three different cases (i) the `classic' case of a star spinning about a principal  axis; (ii) a biaxial star, not spinning about a principal axis; (iii) a triaxial  star spinning steady, but not about a principal axis (as described in Jones, MNRAS {\bf 402}, 2503 (2010)).  The first of these emits only at one frequency; the other two at a pair of harmonically related frequencies.  We show that in all three cases, when written in terms of the original `source parameters', there exist a number of discrete degeneracies, with different parameter values giving rise to the same gravitational wave signal.  We show how these can be removed by suitably restricting the source parameter ranges.  In the case of the model as written down by Jones, there is also a continuous degeneracy.  We show how to remove this  through a suitable rewriting in terms of `waveform parameters', chosen so as to make the specialisations to the other stellar models particularly simple.  We briefly consider the (non-trivial) relation between the assignment of prior probabilities on one set of parameters verses the other.  The results of this paper will be of use when designing strategies for  carrying out searches for such multi-harmonic gravitational wave signals, and when performing parameter estimation in the event of a detection. 
\end{abstract}


\section{Overview} \label{sect:overview}

Rotating neutron stars are potentially detectable sources of continuous gravitational radiation.  They may emit a steady gravitational wave signal because of a non-axisymmetry in their mass distributions, caused either by elastic strains in their solid phase(s), or by magnetic strains sourced by their global magnetic field (see e.g. \citet{aetal_11} for a review).  A star with such a `mountain' will typically emit at a frequency $2f$, where $f$ is the spin frequency.  If the star does not rotate steadily but instead undergoes free precession, the gravitational wave signal is then emitted at frequencies equal (or close to) both $f$ and $2f$ \citep{zs79, ja02}.  However, such precession would normally leave an imprint on the observed radio pulsation received from a pulsar (see e.g. \citet{ja01}).  Such modulations are not typically seen in the known pulsars.  For this reason, most gravitational wave searches to date that have targeted known pulsars have searched only at the frequency $2f$; see \cite{known_pulsars_2014} and references therein.  (The exceptions have been two `narrow band' searches for the Crab pulsar and one for the Vela pulsar, where a small band around $2f$ was searched (\citet{beating_crab_08}, and \citet{LVC_narrowband}).

However, as shown in \citet{jone10}, the presence of a pinned superfluid within the star can change this picture.  A star can then rotate steadily, and still produce gravitational radiation at both $f$ and $2f$, providing the axis about which the pinning takes place does not coincide with a principal axis of the star's moment of inertia tensor.   This motivates the carrying out of searches for such  multi-harmonic signals from  known pulsars, despite their lack of precession.

Given these considerations, we can identify three types of continuous gravitational wave emission.  There is the  general case (as per \citet{jone10}), which we term the \emph{triaxial non-aligned} case, with emission at both $f$ and $2f$.  There is also the simplest case, of a triaxial star spinning about a principal axis.  This is the sort of signal assumed in most targeted gravitational wave searches to-date, and  produces emission at only $2f$.  We term this the \emph{triaxial aligned} case.  There is also an intermediate \emph{biaxial case}, where the star is assumed biaxial.  This produces gravitational radiation at both $f$ and $2f$, with a waveform of intermediate complexity.  The waveform in this case is identical to that of a biaxial precessing star, as considered in the literature \citep{zs79, ja02}, so is an important case to include, but note that in the model of \citet{jone10}, the gravitational wave emission is \emph{not} accompanied by precession. 

A number of parameters  appear in these models; we term these conventional parameter choices the \emph{source parameters}.  It turns out that there are two issues with these parameters, which we address in this paper.  Both issues relate to the existence of degeneracies, i.e. the existence of different values of the source parameters that produce the same detected gravitational wave signal $h(t)$.  Firstly, for all of the models, there exist a number of \emph{discrete} degeneracies.  These are related to discrete symmetries of the star's mass distribution.  Secondly, in the case of the model and parameterisation of \citet{jone10}, there also exists a \emph{continuous} symmetry, i.e. there exists a $1$-parameter family of source parameters that  all produce the same $h(t)$.  

We do two main things in this paper.  We first identify the discrete symmetries of the models written in terms of source parameters.  This allows us to give minimal ranges that are needed in these parameters, removing redundancy in this form of the parameterisation, such that there is a unique set of parameters corresponding to any given star.   We present such ranges in tabular form.   Secondly, we identify a convenient set of   \emph{waveform parameters},   in which the continuous degeneracy present in the parameterisation of \cite{jone10} is removed by effectively reducing the number of parameters by one.  Any future gravitational wave search could be first carried out using this waveform parameterisation.  The conversion to the corresponding $1$-parameter (but possibly more insightful) family of source parameters could then be carried out by making use of formulae given here, with the ranges in source parameters restricted appropriately.  We also (very briefly) discuss the relationship between prior probabilities assigned to the source parameters and the corresponding prior probabilities on the waveform parameters.

The existence of the continuous degeneracy of the model of \citet{jone10} was noted by  \citet{bk14}, who wrote down a reduced parameter set that removed this degeneracy.  The waveform parameters used here are different from the parameter set introduced by \citet{bk14}, and more closely tied to the fundamental scalar quantity, the mass quadrupole moment, that described the gravitational wave emission properties of the star.  Our chosen parameterisation makes the specialisation to simpler forms of gravitational wave emission particularly transparent.

The results of this paper will be of use to gravitational wave observers when devising strategies for carrying out searches for gravitational wave signals with such multiple frequency components.  A study of the issues raised by such searches is currently underway, and will be presented elsewhere (Pitkin et al., in preparation).

The structure of this paper is as follows.  In Section \ref{sect:source_params} we look at the source parameter description, carefully identifying the discrete symmetries, and thereby finding minimal ranges to which the parameters can be restricted.  In Section \ref{sect:waveform_params} we give the waveform parameterisation, in which the continuous degeneracy is removed.  In Section \ref{sect:priors} we briefly discuss the relationship between priors written in terms of the source parameters and priors written in terms of the waveform parameters.  We summarise our findings in Section \ref{sect:summary}.  For the convenience of those carrying out gravitational wave searches, in the Appendix we provide summary tables showing (one possible choice) of ranges appropriate to the source parameters (Table \ref{table:source}) and the waveform parameters (Table \ref{table:waveform}).

\section{Choosing ranges in the source parameters} \label{sect:source_params}

We will consider gravitational wave emission from a star spinning steadily, i.e.  with an angular velocity  $\bf \Omega$, fixed in the inertial frame.  The star has a moment of inertia tensor that is constant in the rotating frame, with principal components $(I_1, I_2, I_3)$.  In the most general case, the rotation axis need not coincide with one of these principal axes; as argued in \citet{jone10}, the presence of an internal pinned superfluid, with pinning axis misaligned with a principal axes, will be of this class.  As described above, and as will be elaborated upon below, there are two special cases, the triaxial aligned case and the biaxial case.  There are a number of features common to all three cases, which we will now describe.  

The signal $h(t)$ received by a gravitational wave detector is given by equation (4) of \cite{jks98}, hereafter JKS:
\begin{equation}
\label{eq:h_of_t}
h(t) = F_+(t) h_+(t) + F_\times(t) h_\times(t) ,
\end{equation}
where the antenna functions $F_+$, $F_\times$ are given by equations (10) and (11) of JKS:
\begin{eqnarray}
\label{eq:F_plus}
F_+(t) &=& \sin\zeta [a(t)\cos 2\psi_{\rm pol} + b(t)\sin 2\psi_{\rm pol}] , \\
\label{eq:F_cross}
F_\times(t) &=& \sin\zeta [b(t)\cos 2\psi_{\rm pol} - a(t)\sin 2\psi_{\rm pol}]  ,
\end{eqnarray}
where $\zeta$ is the angle between the interferometer arms and $\psi_{\rm pol}$ is the polarisation angle; the subscript `pol' has been added to avoid confusion with the Euler angle $\psi$ that appears below.  The functions $a(t)$ and $b(t)$ are complicated functions of source location in the sky (specified by right accession $\alpha$ and declination $\delta$) and detector location on the Earth, as given by equations (12) and (13) of JKS.  The phasing of the signal, as give by JKS equation (14), also depends upon the source location in a complicated way, so we will always need to cover the full sky parameter space of $0 \le \alpha \le 2\pi$, $-\pi/2 \le \delta \le \pi/2$.  Physically, the spin vector can point in any direction,  over the full range in inclination angle $0 < \iota < \pi$ and the full range in polarisation angle  $ 0 \le \psi_{\rm pol} \le 2\pi$.

Collecting these results, we see that in all cases, the following ranges in gravitation wave frequency, sky location and spin vector orientation correspond to physically distinct sources:
\begin{eqnarray}
\label{eq:full_Omega}
0 < & \Omega & < \infty , \\ 
\label{eq:full_alpha}
0 \le & \alpha & \le 2\pi , \\
\label{eq:full_delta}
-\pi/2 \le  & \delta & \le \pi/2 , \\
\label{eq:full_iota}
0 \le & \iota & \le \pi,   \\
\label{eq:full_psi_pol}
0 \le & \psi_{\rm pol} &   < 2\pi .
\end{eqnarray}
Strictly, $\Omega$ is a  function of time, so $\Omega = \Omega(t)$.  In practice, the full phase evolution is often parameterised as a Taylor expansion in $\Omega$ and its time derivatives; by writing the single parameter $\Omega$ we are subsuming all such phase information into the one parameter, to avoid introducing further parameters that have no bearing on our considerations here.

Given the form of equations (\ref{eq:F_plus}) and (\ref{eq:F_cross}), if one only cares about the received waveform $h(t)$,  it is clearly possible to restrict the range in $\psi_{\rm pol}$ further:
\begin{equation}
\label{eq:psi_pol_pi}
0 \le \psi_{\rm pol}  < \pi .
\end{equation}
In fact, as we argue below, if one only cares about the received waveform $h(t)$, and has no `prior' information on the other parameters, it is possible to restrict the range in $\psi_{\rm pol}$ further to  $0 \le \psi_{\rm pol} \le \pi/2$.  However, the full range over a complete circle given in equation (\ref{eq:full_psi_pol}) is the one required to represent all physically distinct stellar configurations, and a value for $\psi_{\rm pol}$ obtained by non-gravitational wave means could lie anywhere in this interval. 

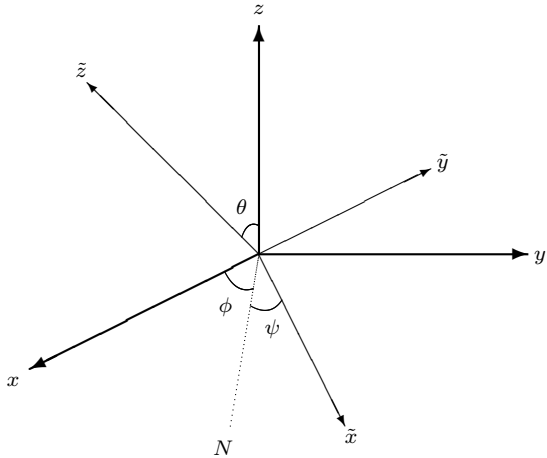
\begin{figure}
\begin{center}
\begin{picture}(100,100)

\thicklines
\put(50,50){\vector(-2,-1){40}}
\put(6,27){$x$}
\put(50,50){\vector(1,0){47}}
\put(98,49){$y$}
\put(50,50){\vector(0,1){40}}
\put(49,92){$z$}

\thinlines
\put(50,50){\vector(1,-2){15}}
\put(65,17){$\tilde x$}
\put(50,50){\vector(2,1){30}}
\put(81,65){$\tilde y$}
\put(50,50){\vector(-1,1){30}}
\put(18,81){$\tilde z$}

\put(42,15){$N$}
\qbezier[50](50,50)(49,44)(45,20)

\qbezier(50, 55)( 48,56)(47,53)
\put(46,57){$\theta$}
\qbezier(44,47)(46,43)(49,44)
\put(43,40){$\phi$}
\qbezier(48.5,41)(52,39)(54,42)
\put(51,36){$\psi$}

\end{picture}
\end{center}
\caption{The orientation of our body is specified by the three standard Euler angles $(\theta, \phi, \psi)$, as labelled above.  The fixed inertial-frame axes are denoted by $(x,y,z)$, while the body-frame axes are denoted by $(\tilde x, \tilde y, \tilde z)$ and rotate about the inertial $z$-axis.  The so-called line of nodes, $N$, lies along the intersection of the $xy$ and $\tilde x \tilde y$ planes.  }
\label{fig:euler_angles}
\end{figure}

We now turn to the form of the polarisation components $h_+(t)$ and $h_\times(t)$.  In the model of  \citet{jone10}, the orientation of the body, and therefore the gravitational wave emission,  depend upon three angles $\theta, \phi, \psi$, basically just Euler angles giving the orientation of the body with respect to the inertial frame; see Figure \ref{fig:euler_angles}.  (This triaxial non-aligned case has the triaxial aligned and biaxial solutions as special cases). 
In the pinned superfluid case $\theta$ and $\psi$ are constants, while $\phi$ is the angle that generates the rotation, so that 
\begin{equation}
\phi = \Omega t + \phi_0 ,
\end{equation}  
with $\phi_0$ a constant, giving the orientation of the body at time $t=0$.  However, as we show below, the phase function that actually appears in the waveforms is $\phi_{\rm gw}$ (or twice this), given by
\begin{equation}
\phi_{\rm gw} = \Omega t + \phi_{{\rm gw},0} ,
\end{equation}
where we have  defined  the constant
\begin{equation}
\label{eq:phi_gw_0}
\phi_{{\rm gw}, 0} \equiv  \phi_0 -\phi_{\rm obs} ,
\end{equation}
where $\phi_{\rm obs}$ is the azimuthal location of the observer.  It was obvious that a constant of this form should appear in the waveform, as it is only the $t=0$ position of the source \emph{relative to the observer} that can affect the received signal.

Also, it is only the asymmetries in the moment of inertia tensor that appear in the waveforms, so we define:
\begin{eqnarray}
\Delta I_{21} &\equiv& I_2 - I_1 , \\
\Delta I_{31} &\equiv& I_3 - I_1 .
\end{eqnarray}

We therefore see that we have a set of ten parameters, which we refer to as the \emph{source parameters}:
\begin{equation}
\label{eq:lambda_source}
{\blambda}_{\rm source} = \{ \Omega, \alpha, \delta, \iota, \psi_{\rm pol}, \Delta I_{21}, \Delta I_{31}, \theta, \phi_{{\rm gw},0}, \psi \} .
\end{equation}
The ranges in the first five parameters that correspond to physically distinct stellar configurations were given in equations (\ref{eq:full_Omega})--(\ref{eq:full_psi_pol}) above.

The ranges and values of the other parameters depends upon which of the three cases we consider, but their maximal ranges are:
\begin{equation}
\label{eq:euler_defaults}
0 \le \theta \le \pi,  \hspace{5mm} 0 \le \phi_{{\rm gw}, 0} \le 2\pi,  \hspace{5mm}  0 \le \psi \le 2\pi ,
\end{equation}
for the Euler-type angles.  For the sake of definiteness, we will also assume there is some restriction on the allowed sizes of the asymmetries in the moment of inertia tensor, although this really depends upon the (poorly constrained) physical mechanisms that produce the deformation in the first place:
\begin{equation} 
\label{eq:Delta_I_defaults}
-\Delta I_{\rm max} < \Delta I_{21} < \Delta I_{\rm max}, \hspace{5mm} -\Delta I_{\rm max} < \Delta I_{31} < \Delta I_{\rm max} .
\end{equation}
Some of these parameters can be set to zero, or simply don't appear, in the triaxial aligned and biaxial waveforms, while the appropriate ranges are also dependent upon the model.

In this Section, our purpose is two-fold:
\begin{enumerate}
\item \label{aim:1} To identify particular ranges in the parameters $\{ \Delta I_{21}, \Delta I_{31}, \theta, \phi_{{\rm gw},0}, \psi \} $ such that all physically distinct configurations of a star's mass quadrupole (which is the quantity responsible for generating the gravitational waves) can be described uniquely.  It will be possible to incorporate all conceivable additional prior information on the parameters (possibly obtained by electromagnetic means) within these ranges.  This will eliminate most of the degeneracies in the waveform.  We provide a summary of a possible choice of source parameter ranges in Table \ref{table:source}.
\item \label{aim:2}To identify and exploit one further discrete symmetry in the waveform connected with the polarisation angle $\psi_{\rm pol}$ that,  in the absence of prior information,  would allow a search to be carried out over a slightly smaller parameter range (leading to a simpler search), together with a rule for generating the other, equally acceptable parameter values, in the event of a successful detection.  
\end{enumerate}

In terms of our first aim, we will exploit the fact that, as we are considering gravitational waves generated by the mass quadruple of the star, we only care about the actual orientation of the star up to a  $\pi$ rotation about any one of the body axes $(\tilde x, \tilde y, \tilde z)$, rendering the full ranges of equation  (\ref{eq:euler_defaults}) redundantly large.  For instance, suppose some glowing hotspot is observed in the rotational equator of a triaxial aligned star, and some astronomer's theory said that this must correspond to the position of the axis of least moment of inertia.  Suppose this hotspot is such that at time $t=0$ it lies on the far side of the star, relative to the observer.  We would then set $\phi_{{\rm gw},0} = \pi$.  However, the $\pi$ rotation symmetry of the mass quadrupole means that there must also be an axis of least moment of inertia on the nearside of the star at $t=0$, so we could equally well set $\phi_{{\rm gw},0} = 0$, i.e. we have the freedom to map all values of $\phi_{{\rm gw},0}$ into the range $0 \le \phi_{{\rm gw},0} <  \pi$.  Other rotational symmetries can be exploited for the biaxial and triaxial non-aligned cases, although their description in terms of Euler angles will be more complicated, as will be described below.  

In terms of our second aim, note that it is obvious from the equations of JKS (i.e.\  equations (\ref{eq:h_of_t})--(\ref{eq:F_cross}) above), a rotation in orientation angle $\psi_{\rm pol} \rightarrow \psi_{\rm pol} + \pi/2$ produces a change in sign of $h(t)$, i.e. $h(t) \rightarrow - h(t)$.  But there is a second way of producing an (also physically distinct) star with waveform $-h(t)$.  Consider emission from a star described by a given set of parameters, producing a waveform $h(t)$.  Now consider emission for a star that has an identical spin vector and sky location, but whose density perturbation $\delta \rho$ away from sphericity is reversed in sign, i.e. $\delta \rho \rightarrow -\delta \rho$.  Such a star will produce a waveform $-h(t)$, and will be described by a different set of parameters.   If follows that if both operations are carried out at once the waveform is unchanged, i.e. there is a degeneracy in the waveform.  Explicitly, given a signal with a particular set of parameters, there will exist three other sets of parameters corresponding to the same signal, obtained, by transforming $\psi_{\rm pol} \rightarrow \psi_{\rm pol} + \pi/2$ and simultaneously transforming some or all of the parameters $\{\theta, \phi_{{\rm gw},0}, \psi, \Delta I_{31}, \Delta I_{21} \}$, in a way that depends upon the choices already made for the allowed ranges of these parameters.  Note that solutions that differ by $\psi_{\rm pol} \rightarrow \psi_{\rm pol} + \pi/2$ will be described by different $\{\theta, \phi_{{\rm gw},0}, \psi, \Delta I_{31}, \Delta I_{21} \}$ parameters, while solutions that differ by $\psi_{\rm pol} \rightarrow \psi_{\rm pol} + \pi$ will be described by the  same values of $\{\theta, \phi_{{\rm gw},0}, \psi, \Delta I_{31}, \Delta I_{21} \}$, so differ only in the orientation of their spin vectors.  The existence and effects of this sort degeneracy were described  long ago in the context of  biaxial precessing stars by \citet{zs79}.  

This means that there several ways in which the polarisation angle can be handled in a gravitational wave search:
\begin{enumerate}
\item If electromagnetic observations have provided a value (or at least small range) in $\psi_{\rm pol}$ (whose value may  lie anywhere in the range $0 < \psi_{\rm pol} < 2\pi$), then this should be used in the search, and a detection will correspond to a single set of parameters $\{ \Delta I_{21}, \Delta I_{31}, \theta, \phi_{{\rm gw},0}, \psi \} $. 
\item If electromagnetic observations do not constrain $\psi_{\rm pol}$, one could search over the interval $0 < \psi_{\rm pol} < 2\pi$, but this would be highly redundant, with a detection potentially manifesting itself at four different points over the searched parameter range, with each inferred parameter set differing by $\pi/2$ in $\psi_{\rm pol}$.
\item More sensibly, one could search over the reduced range $0 < \psi_{\rm pol} < \pi$, but again mindful of a degeneracy, with a detection manifesting itself at two different values of $\psi_{\rm pol}$ within this range, differing by $\pi/2$, and corresponding to two different parameter sets $\{ \Delta I_{21}, \Delta I_{31}, \theta, \phi_{{\rm gw},0}, \psi \} $.  The remaining parameter sets, corresponding to identical waveforms, can then be generated by the simple transformation $\psi_{\rm pol} \rightarrow \psi_{\rm pol} + \pi$, keeping the other parameters fixed.
\item Probably most sensibly of all, one could search over the reduced range $0 < \psi_{\rm pol} < \pi/2$.  This eliminates the discrete degeneracy, and a detection would be expected to manifest itself as a single set of parameters.  In the event of a detection, three other sets of parameters, corresponding to identical waveforms, can then be generated by successive use of the transform $\psi_{\rm pol} \rightarrow \psi_{\rm pol} + \pi/2$, while simultaneously carrying out a transform on (all or some) of the parameter set $\{ \Delta I_{21}, \Delta I_{31}, \theta, \phi_{{\rm gw},0}, \psi \} $, in a way that depends upon the model under consideration, as will be described in this Section.
\end{enumerate}

We will now turn to a consideration of each of our three models.  When we write out the polarisation components $h_+$ and $h_\times$ below, we refer to a gravitational wave detector  with its $1$-arm along $\bf e_\theta$ and its $2$-arm along $\bf e_\phi$, relative to the inertial frame axes $(x,y,z)$ described above.  See \citet{DCC_examples} for the generalisation to an arbitrary detector orientation.

\subsection{Triaxial star, not spinning about a principal axis} \label{sect:tna_source}

In the general case the wave field can be written as (see \citet{jone10} and \citet{DCC_examples} for details): 
\begin{eqnarray}
\label{eq:h_plus_2omega} \nonumber
h_+^{2\Omega} &=&  \frac{2\Omega^2}{r} (1+\cos^2\iota)
\big\{[\Delta I_{21}(\sin^2\psi - \cos^2\psi \cos^2\theta)  \\
& &
- \Delta I_{31}\sin^2\theta]\cos2\phi_{\rm gw}  + \Delta I_{21} \sin 2\psi \cos\theta \sin2\phi_{\rm gw} \big\} , \\
\label{eq:h_cross_2omega} \nonumber
h_\times^{2\Omega} &=&  -  \frac{2\Omega^2}{r} 2\cos\iota
\big\{  \Delta I_{21}\sin2\psi \cos\theta \cos2\phi_{\rm gw}   \\ 
&& \nonumber - [\Delta I_{21} (\sin^2\psi - \cos^2\psi \cos^2\theta)  \\
&& - \Delta I_{31} \sin^2\theta ]  \sin 2\phi_{\rm gw}   \big\} , \\
\label{eq:h_plus_omega} \nonumber
h_+^\Omega &=&  \frac{\Omega^2}{r} \sin\iota \cos\iota
\big\{\Delta I_{21}\sin2\psi \sin\theta \cos\phi_{\rm gw}  \\ 
&& + (\Delta I_{21}\cos^2\psi - \Delta I_{31}) \sin 2\theta \sin\phi_{\rm gw}  \big\} , \\
\label{eq:h_cross_omega} \nonumber
h_\times^\Omega &=&  -  \frac{\Omega^2}{r} \sin\iota
\big\{ (\Delta I_{21}\cos^2\psi - \Delta I_{31}) \sin 2\theta \cos\phi_{\rm gw}  \\ 
&& - \Delta I_{21} \sin 2\psi \sin\theta \sin\phi_{\rm gw}   \big\} . 
\end{eqnarray}
(This differs only trivially from the waveform given in \citet{jone10}, were the observer location was fixed to $\phi_{\rm obs} = -\pi/2$, the  detector with respect to which $h_+$ and $h_\times$ were referred had its 1-arm along  ${\bf e^1} = {\bf e^\phi}$, and its 2-arm along ${\bf e^2} = - {\bf e^\theta}$.  See \citet{DCC_examples} for the equations for the general metric perturbation $h_{ab}$, and for projecting this onto an arbitrarily orientated detector).

The physical and therefore default ranges in the Euler angles are given by equation (\ref{eq:euler_defaults}) above.   One could allow the parameters $\Delta I_{21}$ and  $\Delta I_{31}$ to take either sign (positive or negative).  However, we are free to follow the common convention of rigid body dynamics and choose our axes such that $I_3 > I_2 > I_1$, so that
\begin{equation}
\label{eq:Delta_I_inequality} 
\Delta I_{\rm max} > \Delta I_{31} > \Delta I_{21} > 0 .
\end{equation}

There exist discrete symmetries that can be exploited to reduce the range in parameters further.  The quadrupole moment tensor, and therefore the gravitational wave field, is invariant under rotation of $\pi$ about any one of the three body axes, $O\tilde x$, $O\tilde y$ or $O\tilde z$ .  A rotation about $O\tilde z$ corresponds to $\psi \rightarrow \psi + \pi$ (this is obvious from the definition of $\psi$, but a proof is given in Appendix \ref{sect:pi_about_z}).   The waveform above clearly only depends upon $\sin2\psi$ and $\cos2\psi$ (or as functions that we be re-written in terms of these), so indeed has this symmetry.  This means we can halve the range in this angle to $0 < \psi < \pi$, leaving 
\begin{equation}
0 \le \theta \le \pi, \hspace{5mm} 0 \le \phi_{{\rm gw}, 0} \le 2\pi, \hspace{5mm} 0 \le \psi \le \pi.
\end{equation}

A rotation of $\pi$ about the body's $O\tilde y$ axis corresponds to the mapping $(\theta, \phi_{{\rm gw}, 0},  \psi) \rightarrow (\pi - \theta, \phi_{{\rm gw}, 0} + \pi,  -\psi)$; see Appendix \ref{sect:pi_about_y}.  The waveform above can indeed be shown to posses this symmetry.  This means we can make a further reduction, halving the parameter range of any one (but only one) of the parameters $(\theta, \phi_{{\rm gw}, 0},  \psi)$.   (Note that the waveform must also be invariant under a $\pi$ rotation about $O\tilde x$, but this is equivalent to the composition of the above two rotations, so cannot generate any further reduction in the parameter space.)  This means there are three options for the Euler angle parameters:
\begin{eqnarray}
\label{eq:tna_source_preferred}
&0& \le \theta \le \pi/2, \hspace{3mm}   0 \le \phi_{{\rm gw}, 0} \le 2\pi, \hspace{3mm}    0 \le \psi \le \pi, \hspace{3mm}  {\bf or} \\
&0& \le \theta \le \pi, \hspace{3mm}   0 \le \phi_{{\rm gw}, 0} \le \pi, \hspace{3mm}   0 \le \psi \le \pi,  \hspace{3mm}    {\bf or} \\
&0& \le \theta \le \pi, \hspace{3mm}   0 \le \phi_{{\rm gw}, 0} \le 2\pi, \hspace{3mm}  0 \le \psi \le \pi/2 .
\end{eqnarray}
Any one of these choices, together with the ranges of equation (\ref{eq:Delta_I_inequality}) and the parameter ranges of equations (\ref{eq:full_Omega})--(\ref{eq:full_psi_pol}),  will always be able to accommodate any additional prior information on the star.

However, this parameter space is redundantly large from the point of view of carrying out a gravitational wave search without additional prior information.  As argued above, if there exists a star producing a wavefield with components $h^{\Omega, 2\Omega}_{+, \times}$ there must exist another, physically distinct star, whose wavefield has the sign of  all these components reversed, i.e. $h^{\Omega, 2\Omega}_{+, \times} \rightarrow - h^{\Omega, 2\Omega}_{+, \times}$.  Physically, this corresponds to reversing the sign of the density perturbation $\delta \rho$ that deforms the star away from spherical symmetry.  The transformation that produces the mapping  $h^{\Omega, 2\Omega}_{+, \times} \rightarrow - h^{\Omega, 2\Omega}_{+, \times}$ is rather complicated.  The details are given in Appendix \ref{sect:h_to_minus_h}, and involve a transformation mixing the Euler angles $(\theta, \phi_{\rm gw,0}, \psi)$ and also mixing the amplitude parameters $(\Delta I_{21}, \Delta I_{31})$.   If this transformation is carried out together  with the operation $\psi_{\rm pol} \rightarrow \psi_{\rm pol} + \pi/2$, the full waveform $h(t)$ is invariant.   The significance of this is that a further reduction in the parameter space is possible.  One option is to reduce the range in $\psi_{\rm pol}$ to the range:
\begin{equation}
0 < \psi_{\rm pol} < \pi/2  .
\end{equation}
There is presumably another  option, involving some reduction in the parameter set $(\theta, \phi_{\rm gw,0}, \psi, \Delta I_{21}, \Delta I_{31})$, but it is not clear how to implement this second option.  For instance, the Euler angles aren't simply increased by $\pi$ or multiplied by $-1$, so the reduction presumably isn't a simple halving their ranges (see equations (\ref{eq:euler_transform})--(\ref{eq:Delta_I_31_transform}) in Section \ref{sect:h_to_minus_h}).  

If a gravitational wave search is carried out over this restricted range $0 < \psi_{\rm pol} < \pi/2 $, three other equally acceptable solutions can be obtained by successive applications of the transformation 
\begin{equation}
\label{eq:transform_tna1_source}
\psi_{\rm pol} \rightarrow \psi_{\rm pol} + \pi/2 ,
\end{equation}
with a corresponding transformation of $(\theta, \phi_{\rm gw,0}, \psi, \Delta I_{21}, \Delta I_{31})$ given by equations (\ref{eq:euler_transform})--(\ref{eq:Delta_I_31_transform}).   (If one wishes the parameters to remain confined to one of the minimal ranges identified above, further transformations may be needed, e.g.\ the $\pi$ rotation about $O\tilde z$ of Appendix  \ref{sect:pi_about_z}).

Note that if we weren't enforcing the inequalities of equation (\ref{eq:Delta_I_inequality}) and were instead allowing the quantities $\Delta I_{21}$ and $\Delta I_{31}$ to take either sign, the operation $\delta \rho \rightarrow -\delta \rho$ could be achieved much more simply, by making the replacements $\Delta I_{21} \rightarrow - \Delta I_{21}$ and $\Delta I_{31} \rightarrow - \Delta I_{31}$.  This would, however, lead to different and less easily derivable choices in minimal ranges of the Euler-type angles.

See the first column of Table \ref{table:source} for a summary of a possible choice of parameter ranges, where, for definiteness,   we use the ranges of equation (\ref{eq:tna_source_preferred}) for the Euler-like angles.

\subsection{Intermediate case: A biaxial star, not spinning about a principal axis} \label{sect:biax_source}

The simplest (and conventional way) of describing a biaxial body is to single out the $\tilde z$-axis as special, i.e. to set $I_1 = I_2$, so that $\Delta I_{21} = 0$.  The waveform can then be shown to be \citep{DCC_examples}
\begin{eqnarray}
h_+^{2\Omega} &=&  -  \frac{2\Omega^2}{r} (1+\cos^2\iota) \Delta I_{31} 
\sin^2\theta \cos2\phi_{\rm gw} ,\\
\label{eq:h_cross_biaxial}
h_\times^{2\Omega} &=& -  \frac{2\Omega^2}{r} 2\cos\iota \Delta I_{31} 
\sin^2\theta \sin2\phi_{\rm gw} , \\
\label{eq:h_plus_biaxial}
h_+^\Omega &=&  -  \frac{\Omega^2}{r} \sin\iota \cos\iota \Delta I_{31} 
\sin 2\theta \sin\phi_{\rm gw} ,\\
h_\times^\Omega &=&  \frac{\Omega^2}{r} \sin\iota \Delta I_{31} 
\sin 2\theta \cos\phi_{\rm gw} .
\end{eqnarray}
This is the wave field of a biaxial star spinning about an axis other than a principal axis.  In the model of \cite{jone10} it corresponds to a non-precessing star with a pinned superfluid.  It is also identical to the GW field of a precessing biaxial star without pinning, of the sort considered by \cite{zs79} and JKS.  So, it is an important case to cover.

By singling out the $\tilde z$-axis as the symmetry axis, we have $\Delta I_{21} = 0$.  Having made this choice, we can no longer insist that $I_3$ is the axis of greatest moment of inertia.  Instead, we must allow for the star being either oblate ($\Delta I_{31} > 0$) or  prolate ($\Delta I_{31} < 0$), depending upon the sign of $\Delta I_{31}$, so we have:
\begin{equation}
\label{eq:Delta_for_biaxial}
\Delta I_{21} = 0, \hspace{5mm} -\Delta I_{\rm max} < \Delta I_{31} < \Delta I_{\rm max} .
\end{equation}

Now consider the Euler angles, whose `default' ranges were given in equation (\ref{eq:euler_defaults}).   We can again exploit symmetries.  The waveform no longer depends upon the angle $\psi$, a consequence of the axisymmetry of the body about the $O\tilde z$ body axis, so this angle is removed from our considerations.  As described in Section \ref{sect:pi_about_y} the operation of performing a rotation of  $\pi$  rotation about the $O\tilde y$ axis takes the form  $(\theta, \phi_{{\rm gw}, 0}) \rightarrow (\pi - \theta, \phi_{{\rm gw}, 0} + \pi)$, allowing us to halve the range in $\theta$ or $\phi_{{\rm gw}, 0}$.  We can therefore have:
\begin{eqnarray}
\label{eq:biaxial_source_choice_1}
&& 0 \le \theta \le \pi/2 ,  \hspace{5mm}    0 \le \phi_{{\rm gw}, 0} \le 2\pi,  \hspace{5mm}   {\bf or} \\
\label{eq:biaxial_source_choice_2}
&& 0 \le \theta \le \pi , \hspace{5mm}  0 \le \phi_{{\rm gw}, 0} \le \pi .
\end{eqnarray}
Either one of these two options, together with the parameter ranges of equation (\ref{eq:Delta_for_biaxial}) and equations (\ref{eq:full_Omega})--(\ref{eq:full_psi_pol}),  will always be able to accommodate any additional prior information on the star.  

However, this parameter space is again redundantly large from the point of view of carrying out a gravitational wave search without addition prior information.  To see this, note that the transformation  $\Delta I_{31} \rightarrow -\Delta I_{31}$ (equivalent to the transformation $\delta \rho \rightarrow -\delta \rho$) changes the sign of the above polarisation components, i.e. $h^{\Omega, 2\Omega}_{+, \times} \rightarrow - h^{\Omega, 2\Omega}_{+, \times}$, thereby flipping the sign of $h(t)$.  The transformation $\psi_{\rm pol} \rightarrow \psi_{\rm pol} + \pi/2$ also flips the sign of $h(t)$, so the two transformations together leave $h(t)$ unchanged.  It follows we can reduce the range in $\Delta I_{31}$ or the range in $\psi_{\rm pol}$, i.e.\ we have the choice
\begin{eqnarray}
&& 0 < \psi_{\rm pol} < \pi/2,  \hspace{5mm}  -I_{\rm max} < \Delta I_{31} < I_{\rm max},   \hspace{5mm}  {\bf or} \\
&& 0 < \psi_{\rm pol} < \pi,  \hspace{5mm}   0 < \Delta I_{31} < I_{\rm max} .
\end{eqnarray}
This degeneracy was noted by \citet{zs79}.  A gravitational wave search for a biaxial star could then be carried out with either of the two choices hardwired in, with the understanding that in the event of a detection, three other equally valid solution can be obtained via successive uses of the transformation: 
\begin{equation}
\label{eq:transform_biaxial_source}
\psi_{\rm pol} \rightarrow \psi_{\rm pol} + \pi/2, \hspace{5mm} \Delta I_{31} \rightarrow -\Delta I_{31} .
\end{equation}

See the middle column of Table \ref{table:source} for a summary of a possible choice in parameter ranges, where for definiteness we choose the ranges  of equation (\ref{eq:biaxial_source_choice_1}) for the Euler-like angles.

\subsection{Simplest case: A triaxial star, spinning about a principal axis}      \label{sect:ta_source}

This is the standard case of a triaxial star spinning about a principal axis, emitting only at $2\Omega$, i.e. the sort of emission normally assumed in continuous gravitational wave searches.  The convention is to choose the rotation axis to be the $z$-axis.  This is accomplished by setting $\theta = 0$ in equations (\ref{eq:h_plus_2omega}--\ref{eq:h_cross_omega})   to give
\begin{eqnarray}
\label{eq:h_plus_triaxial_GW}
h_+ &=&  -  \frac{2\Omega^2}{r} \Delta I_{21} (1+\cos^2 \iota) \cos 2[\Omega t + (\phi_{{\rm gw},0}+\psi)] , \\
\label{eq:h_cross_triaxial_GW}
h_\times &=&  -   \frac{2\Omega^2}{r} \Delta I_{21} 2\cos \iota \sin  2[\Omega t + (\phi_{{\rm gw},0} +\psi)] .
\end{eqnarray}
Note that the parameters angles $\phi_{{\rm gw}, 0}$ and $\psi$ are degenerate, i.e. only their sum appears in the waveform.

We are free to lay down the $(O\tilde x, O\tilde y)$ axes such that $\Delta I_{21} > 0$.   The symmetry of rotating by $\pi$ about $O\tilde z$ then corresponds to $(\phi_{{\rm gw}, 0} + \psi) \rightarrow (\phi_{{\rm gw}, 0} + \psi) + \pi$.  The waveform clearly has this symmetry,  suggesting we need to cover the range $0 < (\phi_{{\rm gw}, 0} + \psi)  < \pi$.   As we have fixed $\theta = 0$, this rotation about $O\tilde z$ is the only angular degree of freedom, so there are no further symmetries we can exploit, corresponding to rotations about $O\tilde x$ or $O\tilde y$.  So, for a triaxial aligned rotator, the set of physically distinct configurations is spanned by:
\begin{equation}
\label{eq:ta_DI_and_phi_ranges}
\Delta I_{21} > 0, \hspace{5mm} 0 \le  (\phi_{{\rm gw},0}+\psi) < \pi .
\end{equation}
These choices, together with the parameter ranges of equations (\ref{eq:full_Omega})--(\ref{eq:full_psi_pol}),  will always be able to accommodate any additional prior information on the star.

However, these parameter ranges are redundantly large, in the sense that, in the absence of such prior information, smaller parameter ranges can be used to carry out gravitational wave searches.  To see this, note that the waveform changes sign ($h(t) \rightarrow -h(t)$) when either one of the following transformations is performed:
\begin{enumerate}
\item $\psi_{\rm pol} \rightarrow \psi_{\rm pol} + \pi/2 $ ,
\item $(\phi_{{\rm gw}, 0} + \psi) \rightarrow (\phi_{{\rm gw}, 0} + \psi) + \pi/2$ .
\end{enumerate}
The second of these transformation is equivalent to swapping over the  axes of largest and smallest moment of inertia that lie in the rotational equatorial plane; as such it is not the same as the transformation $\delta \rho \rightarrow -\delta \rho$ discussed above, but rather flips the sign of $h(t)$ in a way that preserves  our choice of fixing the $O\tilde z$ axis as the rotation axis.

It follows that, in carrying out a gravitational wave search,  we can reduce the range in any one (but only one) of these parameters, so the options are:
\begin{eqnarray}
 0 \le \psi_{\rm pol} \le \pi/2,  &&   0 \le (\phi_{{\rm gw}, 0} + \psi) \le \pi,   \hspace{5mm}  {\bf or} \\
 0 \le \psi_{\rm pol} \le \pi, &&    0 \le (\phi_{{\rm gw}, 0} + \psi) \le \pi/2 .
\end{eqnarray}
The first of these three options has traditionally been used in gravitational wave searches (see e.g. \citet{known_pulsars_2014}), although it should be noted that  the phase angle that appears in the literature is  actually
\begin{equation}
\Phi_{{\rm GW}, 0} = 2(\phi_{{\rm gw}, 0} + \psi) ,
\end{equation}
so that searches have  traditionally searched over the range $0 < \Phi_{{\rm GW}, 0} < 2\pi$.

If one of these restricted parameter spaces is used and a signal detected with parameters $(\psi_{\rm pol}, (\phi_{{\rm gw},0}+\psi))$, three additional solutions can be obtained by successive applications  of the transform
\begin{equation}
\label{eq:transform_ta_source}
\psi_{\rm pol} \rightarrow \psi_{\rm pol} + \pi/2 , \hspace{5mm} 
(\phi_{{\rm gw}, 0} + \psi) \rightarrow (\phi_{{\rm gw}, 0} + \psi) + \pi/2 .
\end{equation}
Without additional (non-gravitational wave information) all such solutions are equally valid.  (If one wishes the parameters to remain confined to, say, the range $0 \le (\phi_{{\rm gw}, 0} + \psi) \le \pi$, then a further transformation $(\phi_{{\rm gw}, 0} + \psi) \rightarrow (\phi_{{\rm gw}, 0} + \psi) + \pi$ can be applied when necessary).

See the final column of Table \ref{table:source} for a summary of these possible choices in parameter ranges.

\section{Reformulating in terms of waveform parameters} \label{sect:waveform_params}

The $10$-parameter triaxial non-aligned waveform of Section  \ref{sect:tna_source} contains a degeneracy, and in fact only depends upon 9 parameters.   An easy way of seeing this is to note that if the cosine and sine terms in each of the four equations giving the polarisation components (equations (\ref{eq:h_plus_2omega})--(\ref{eq:h_cross_omega})) are combined into single trigonometric terms (essentially writing the equations  in `amplitude-phase' form), the five parameters $(\theta, \phi_{{\rm gw},0}, \psi, \Delta I_{21}, \Delta I_{31})$   appear in only four different combinations.  Another, possibly more insightful, way of understanding this is to return to first principles, making use of the multipole formalism for gravitational wave emission, as described in \cite{thor80}.  

The fundamental quantities that appears in  the wave generation equations Thorne are the mass quadrupole moment scalars, related to the source's density field $\rho$ by equation (5.27a) of \cite{thor80}:
\begin{equation}
\label{eq:I_2m}
I^{2m} = \frac{16\pi\sqrt{3}}{15} \int \rho Y_{2m}^* r^2 \, dV .
\end{equation}
The transverse traceless (TT) description of the GW field is given by equation  (4.3) of \cite{thor80}:
\begin{equation}
\label{eq:h_ab_TT}
h_{ab}^{\rm TT}(t) = \frac{1}{r} \sum_m \ddot I^{2m} T_{ab}^{{\rm E}2,2m} ,
\end{equation}
where $T_{ab}^{{\rm E}2,2m}$ is a tensor spherical harmonic.  

For rigid rotation about the $z$-axes at rate $\Omega$, the mass quadrupole scalars can be shown to take the from \citep{DCC_examples} 
\begin{equation}
I^{2m} = C_{2m}^{\rm complex} e^{-im(\Omega t + \phi_0)} ,
\end{equation}
where $C^{\rm complex}_{2m}$ is a complex number that encodes details of the source, and $\phi_0$ is as defined above, i.e. a phase angle giving the rotational phase of the body at time $t=0$.  We can write $C_{2m}^{\rm complex}$ in amplitude-angle form:
\begin{equation}
C_{2m}^{\rm complex} = C_{2m} e^{i\Phi_{2m}} ,
\end{equation}
where $C_{2m} \equiv |C_{2m}^{\rm complex}| \ge 0$ and $0 < \Phi_{2m} < 2\pi$.  

The waveform for an arbitrary rigidly rotating source is then given by equation (\ref{eq:h_ab_TT}).  In writing it down, it is convenient to include some additional factors in our amplitude parameters; we define
\begin{equation}
\tilde C_{2m} = \frac{\Omega^2}{r} \sqrt{\frac{5}{2\pi}} C_{2m} ,
\end{equation}
so that the waveform can then be shown to  take the very simple form \citep{DCC_examples}:
\begin{eqnarray}
h^+(2\Omega) &=& 
- \tilde C_{22} \cos[2\Omega t + \Phi^C_{22}] (1+\cos^2\iota) , \\
h^\times(2\Omega) &=& 
- \tilde C_{22} \sin[2\Omega t + \Phi^C_{22}] 2\cos\iota , \\
h^+(\Omega) &=& 
-  \frac{1}{2} \tilde C_{21} \cos[\Omega t + \Phi^C_{21}] \sin\iota \cos\iota , \\
h^\times(\Omega) &=& 
-  \frac{1}{2} \tilde C_{21} \sin[\Omega t + \Phi^C_{21}] \sin\iota ,
\end{eqnarray}
where the phases $\Phi^C_{2m}$ are related to previously introduced quantities  by
\begin{eqnarray}
\label{eqn:Phi_22_relation}
\Phi^C_{22} &=& 2 \phi_{{\rm gw},0}  - \Phi_{22} ,\\
\label{eqn:Phi_21_relation}
\Phi^C_{21} &=& \phi_{{\rm gw},0} - \Phi_{21} .
\end{eqnarray}
These equations can then be specialised to the three cases considered above.  They are clearly rather simple in form, with all the (potentially) complicated details of the source parameters being buried within the amplitudes $\tilde C_{22}$ and $\tilde C_{21}$, and the phases $\Phi^C_{22}$ and $\Phi^C_{21}$.  

This approach also has the advantage of making the counting of the number of parameters more straightforward.  We can count as follows.  We need the set of five parameters 
$ \{ \Omega, \alpha, \delta, \iota, \psi_{\rm pol} \}$ giving the spin frequency, sky location, and spin orientation of the source, as before.  (The maximal ranges in these parameters were given in equations (\ref{eq:full_Omega})--(\ref{eq:full_psi_pol}) earlier).  For a steadily rotating source emitting gravitational waves only at $2\Omega$, we then have the amplitude-phase pair $\tilde C_{22}, \Phi^C_{22}$ also, giving seven parameters, consistent with the number of source parameters in this case.  However, in the triaxial non-aligned case, where the  $\Omega$-harmonic is present too, we also have the amplitude-phase pair $\tilde C_{21}, \Phi^C_{21}$.  This gives a total of nine parameters, not the ten that one would arrive at by examining the waveform as written previously, confirming the existence of a continuous degeneracy in the source parameters in this triaxial non-aligned case.  For the biaxial case, we will find that there is a particular relation between the phases $ \Phi^{\rm C}_{21}$ and $ \Phi^{\rm C}_{22}$, giving eight parameters, equal to the number of source parameters, so there is no degeneracy for biaxial stars, only for triaxial non-aligned ones.

We can collect the relevant parameters  together to give the nine waveform parameters:
\begin{equation}
\label{eq:lambda_waveform}
{\blambda}_{\rm waveform} = \{ \Omega, \alpha, \delta, \iota, \psi_{\rm pol}, \tilde C_{21},   \Phi^{\rm C}_{21},  \tilde C_{22},   \Phi^{\rm C}_{22} \} .
\end{equation}
Comparing with the ten source parameters of equation (\ref{eq:lambda_source}), we see that  the first five parameters $\{ \Omega, \alpha, \delta, \iota, \psi_{\rm pol} \}$ are common between the two parameterisations, while the set of five source parameters $\{  \Delta I_{21}, \Delta I_{31}, \theta, \phi_{{\rm gw},0}, \psi  \}$ are replaced by the set of four waveform parameters $\{ \tilde C_{21},   \Phi^{\rm C}_{21},  \tilde C_{22},   \Phi^{\rm C}_{22} \}$. 

It would therefore seem that there may be an advantage in using the waveform parameters, rather than the source parameters that naturally come out of rigid body calculations.  Let us look at the waveform parameter description of the gravitational wave signal for the three particular cases of interest.  We have two goals: (i) to relate the source parameters to the waveform parameters, and (ii) to identify sensible ranges to search over in the waveform parameters.   A summary of the identified parameter ranges is given in the Appendix; see table \ref{table:waveform}.

\subsection{Triaxial star, not spinning about a principal axis} \label{sect:tna_waveform}

Starting with equation (\ref{eq:I_2m}), the motion of a triaxial non-aligned star leads to \citep{DCC_examples}
\begin{eqnarray}
\nonumber
I^{22} &=&  -  e^{-2i(\Omega t + \phi_0)} \sqrt{\frac{8\pi}{5}}
[\Delta I_{21}(\sin^2\psi-\cos^2\psi\cos^2\theta)- \\
\label{eq:I_22_TNA} & & \hspace{15mm} 
\Delta I_{31}\sin^2\theta + i\Delta I_{21}\sin2\psi\cos\theta] , \\
\nonumber
I^{21} &=&  - e^{-i(\Omega t + \phi_0)} \sqrt{\frac{8\pi}{5}}
[\Delta I_{21}\sin2\psi\sin\theta + \\
\label{eq:I_21_TNA} & & \hspace{15mm} 
i(\Delta I_{21}\cos^2\psi - \Delta I_{31})\sin2\theta] ,
\end{eqnarray}
so that
\begin{eqnarray}
\nonumber
C^{\rm complex}_{22} &=&  - \sqrt{\frac{8\pi}{5}}
[\Delta I_{21}(\sin^2\psi-\cos^2\psi\cos^2\theta) \\
& & \label{eq:C_22_complex} \hspace{5mm} -\Delta I_{31}\sin^2\theta + i\Delta I_{21}\sin2\psi\cos\theta] ,\\
\nonumber
C^{\rm complex}_{21} &=&   - \sqrt{\frac{8\pi}{5}}
[\Delta I_{21}\sin2\psi\sin\theta + \\
& & \label{eq:C_21_complex} \hspace{5mm} i(\Delta I_{21}\cos^2\psi - \Delta I_{31})\sin2\theta] ,
\end{eqnarray}
from which we see
\begin{eqnarray}
\nonumber
\tilde C_{22} &=&  \frac{\Omega^2}{r} 2
\{[\Delta I_{21}(\sin^2\psi-\cos^2\psi\cos^2\theta)-  \\
\label{eq:C_22_source_params} & &   \Delta I_{31}\sin^2\theta]^2 + (\Delta I_{21}\sin2\psi\cos\theta)^2\}^{1/2} ,\\
\nonumber
\tilde C_{21} &=&   \frac{\Omega^2}{r} 2 
\{(\Delta I_{21}\sin2\psi\sin\theta)^2 + \\
\label{eq:C_21_source_params} & & (\Delta I_{21}\cos^2\psi - \Delta I_{31})^2\sin^2 2\theta\}^{1/2} ,\\
\nonumber
\Phi^C_{22} &=& 2\phi_{{\rm gw},0}  \\
\label{eq:Phi_22_source_params}  &-& \tan^{-1} 
\frac{\Delta I_{21}\sin2\psi\cos\theta}{\Delta I_{21}(\sin^2\psi-\cos^2\psi\cos^2\theta)-\Delta I_{31}\sin^2\theta} ,\\
\label{eq:Phi_21_source_params}
\Phi^C_{21} &=& \phi_{{\rm gw},0} - \tan^{-1} 
\frac{(\Delta I_{21}\cos^2\psi - \Delta I_{31})\sin2\theta}{\Delta I_{21}\sin2\psi\sin\theta} .
\end{eqnarray}
If values are given for the quantities $\{ \tilde C_{22}, \Phi^C_{22}, \tilde C_{21}, \Phi^C_{21} \}$, as would be the case in the event of a detection, the above four equations in the five unknowns $\{ \Omega^2\Delta I_{21}/r, \Omega^2\Delta I_{31}/r, \theta, \psi, \phi_{{\rm gw},0} \}$ would then generate a $1$-parameter family of solutions.  Note that, when evaluating the inverse tangent functions of equations (\ref{eq:Phi_22_source_params}) and  (\ref{eq:Phi_21_source_params}), care must be taken to select the correct root so as to correctly reconstruct the complex mass numbers  of equations  (\ref{eq:C_22_complex}) and  (\ref{eq:C_21_complex}).

Having found the algebraic relationship between the source and waveform parameters we can now turn to the issue of selecting \emph{ranges} in the waveform parameters.  Careful study of equations (\ref{eq:C_22_source_params}) and  (\ref{eq:C_21_source_params}) shows that if one selects $\Delta I_{21}$ and $\Delta I_{31}$ according to equation (\ref{eq:Delta_I_inequality}) then the corresponding bounds on the amplitude parameters are:
\begin{eqnarray}
0 &\le& \tilde C_{22} \le  \frac{2\Omega^2}{r} \Delta I_{\rm max} , \\
0 &\le& \tilde C_{21} \le  \frac{2\Omega^2}{r}  \Delta I_{\rm max} .
\end{eqnarray}
For the two phase parameters $\Phi^C_{2m}$, the default range is: 
\begin{equation}
0 < \Phi^C_{2m} < 2\pi .
\end{equation}
Together with the ranges given in equations (\ref{eq:full_Omega})--(\ref{eq:full_psi_pol}), these ranges are sufficiently wide to cover all physically distinct stellar configurations, and accommodate all possible additional information obtained by non-gravitational wave means.

However, from the point of view of carrying out a gravitational wave search without such extra information, these ranges are redundantly large.  The polarisation components change sign under the operation $\Phi^C_{2m} \rightarrow \Phi^C_{2m} + \pi$.  The waveform $h(t)$ also changes sign under the operation $\psi_{\rm pol} \rightarrow \psi_{\rm pol} + \pi/2$, so we can halve the range in one or other of the polarisation angle or the phases.  We therefore have the options:
\begin{eqnarray}
&&0 < \psi_{\rm pol} < \pi/2, \hspace{5mm} 0 < \Phi^C_{2m} < 2\pi, \hspace{5mm}  {\bf or} \\ 
&& 0 < \psi_{\rm pol} < \pi,   \hspace{9mm}  0  < \Phi^C_{2m} < \pi .
\end{eqnarray}
If one or other of these restricted ranges are employed in  a search, and a detection is made with parameters $(\psi_{\rm pol}, \Phi^{\rm C}_{2m})$, three other equally acceptable solutions can be obtained through successive applications of the transformation
\begin{equation}
\psi_{\rm pol} \rightarrow \psi_{\rm pol} + \pi/2, \hspace{5mm} \Phi^C_{2m} \rightarrow \Phi^C_{2m} + \pi .
\end{equation}

See the first column of Table \ref{table:waveform} for a summary of these possible choices in parameter ranges.

\subsection{Intermediate case: A biaxial star, not spinning about a principal axis}  \label{sect:biax_waveform}

Setting $\Delta I_{21}=0$ in equations (\ref{eq:I_22_TNA}) and (\ref{eq:I_21_TNA}) leads to
\begin{eqnarray}
I^{22} &=&   e^{-2i(\Omega t + \phi_0)} \sqrt{\frac{8\pi}{5}} \Delta I_{31} \sin^2\theta ,  \\
I^{21} &=&    i e^{-i(\Omega t + \phi_0)} \sqrt{\frac{8\pi}{5}} \Delta I_{31} \sin2\theta  ,
\end{eqnarray}
so that
\begin{eqnarray}
\label{eq:C_22_biaxial}
C^{\rm complex}_{22} &=& \sqrt{\frac{8\pi}{5}} \Delta I_{31} \sin^2\theta    ,\\
\label{eq:C_21_biaxial}
C^{\rm complex}_{21} &=& \sqrt{\frac{8\pi}{5}} \Delta I_{31} \sin2\theta e^{ i \pi/2} .
\end{eqnarray}
As discussed in Section \ref{sect:biax_source}, we are always free to insist $0 \le \theta \le \pi/2$, as made explicit in equation (\ref{eq:biaxial_source_choice_1}).  With this choice, both $\sin^2\theta$ and $\sin 2\theta$ will always be non-negative.  In contrast, $\Delta I_{31}$ can be either positive or negative, so we should treat the $\Delta I_{31} > 0$ and $\Delta I_{31} < 0$ cases separately.

For $\Delta I_{31} > 0$ case, we can read-off
\begin{eqnarray}
\label{eq:C_22_positive}
C_{22} &=&  \sqrt{\frac{8\pi}{5}} \Delta I_{31} \sin^2\theta, \hspace{5mm} \Phi_{22} = 0 , \\
\label{eq:C_21_positive}
C_{21} &=&  \sqrt{\frac{8\pi}{5}} \Delta I_{31} \sin2\theta, \hspace{5mm} \Phi_{21} = \pi/2 .
\end{eqnarray}
Converting to the parameters $\tilde C_{2m}$ and $\Phi^{\rm c}_{2m}$ that actually appear in the waveform:
\begin{eqnarray}
\label{eq:oblate_params_2}
\tilde C_{22} &=&  \frac{2\Omega^2}{r} \Delta I_{31} \sin^2\theta, \hspace{5mm} \Phi^{\rm c}_{22} = 2\phi_{{\rm gw},0} , \\
\label{eq:oblate_params_1}
\tilde C_{21} &=&  \frac{2\Omega^2}{r} \Delta I_{31} \sin2\theta, \hspace{5mm} \Phi^{\rm c}_{21} = \phi_{{\rm gw},0} - \frac{\pi}{2} .
\end{eqnarray}
These equations can be inverted to give:  
\begin{eqnarray}
\frac{\Omega^2 \Delta I_{31}}{r}  &=&  \frac{1}{2} \tilde C_{22}  \left[1+\left(\frac{\tilde C_{21}}{2\tilde C_{22}}\right)^2\right] ,\\
\phi_{{\rm gw},0} &=& \Phi^C_{21} + \frac{\pi}{2}   , \\
\label{eq:tan_theta_biax}
\tan \theta &=& \frac{2 \tilde C_{22}}{ \tilde C_{21} } .
\end{eqnarray}
Note that in this case 
\begin{equation}
\label{eq:Phi_relation_oblate}
\Phi^{\rm c}_{22} = 2\Phi^{\rm c}_{21} + \pi ,
\end{equation}
a relation that could be hardwired into any search for oblate biaxial stars using the waveform parameterisation.

For $\Delta I_{31} < 0$ case, we can read-off
\begin{eqnarray}
C_{22} &=&  -\sqrt{\frac{8\pi}{5}} \Delta I_{31} \sin^2\theta, \hspace{5mm} \Phi_{22} = \pi , \\
C_{21} &=&  -\sqrt{\frac{8\pi}{5}} \Delta I_{31} \sin2\theta, \hspace{5mm} \Phi_{21} = -\pi/2 .
\end{eqnarray}
Converting to the parameters $\tilde C_{2m}$ and $\Phi^{\rm c}_{2m}$ that actually appear in the waveform:
\begin{eqnarray}
\label{eq:prolate_params_2}
\tilde C_{22} &=&  -\frac{2\Omega^2}{r} \Delta I_{31} \sin^2\theta, \hspace{5mm} \Phi^{\rm c}_{22} = 2\phi_{{\rm gw},0} + \pi , \\
\label{eq:prolate_params_1}
\tilde C_{21} &=&  -\frac{2\Omega^2}{r} \Delta I_{31} \sin2\theta, \hspace{5mm} \Phi^{\rm c}_{21} = \phi_{{\rm gw},0} + \frac{\pi}{2} .
\end{eqnarray}
These equations can be inverted to give:  
\begin{eqnarray}
\frac{\Omega^2 \Delta I_{31}}{r}  &=&  -\frac{1}{2} \tilde C_{22}  \left[1+\left(\frac{\tilde C_{21}}{2\tilde C_{22}}\right)^2\right] ,\\
\phi_{{\rm gw},0} &=& \Phi^C_{21}  - \frac{\pi}{2}   ,
\end{eqnarray}
with $\theta$ given by equation (\ref{eq:tan_theta_biax}).  Note that in this case 
\begin{equation}
\label{eq:Phi_relation_prolate}
\Phi^{\rm c}_{22} = 2\Phi^{\rm c}_{21}  ,
\end{equation}
a relation that could be hardwired into any search for prolate biaxial stars using the waveform parameterisation.  In the event of a successful detection, the measured values for the waveform parameters could then be inserted into the equations above, to deduce the corresponding source parameters.

To identify the ranges in these waveform parameters to search over, we can convert the ranges in the source parameters of equations (\ref{eq:Delta_for_biaxial}) and (\ref{eq:biaxial_source_choice_1})  using equations  (\ref{eq:oblate_params_2})--(\ref{eq:oblate_params_1}) above (or, equivalently, equations  (\ref{eq:prolate_params_2})--(\ref{eq:prolate_params_1})), to give:
\begin{equation}
0 < \tilde C_{2m} < \frac{2\Omega^2}{r} \Delta I_{\rm max},
\end{equation}
\begin{equation}
0 \le \Phi^{\rm C}_{21} < 2\pi ,
\end{equation}
being mindful to use \emph{both} equations (\ref{eq:Phi_relation_oblate}) (for oblate stars) and equation (\ref{eq:Phi_relation_prolate}) (for prolate stars) to calculate $\Phi^{\rm C}_{22}$ as a function of $\Phi^{\rm C}_{21}$.
 Together with the parameter ranges of equations (\ref{eq:full_Omega})--(\ref{eq:full_psi_pol}), these ranges are wide enough to accommodate all physically distinct stellar  configurations.

However, from the point of view of carrying out a gravitational wave search, these ranges are redundantly large.  The waveform changes sign under the operation $\psi_{\rm pol} \rightarrow \psi_{\rm pol}  + \pi/2$.  It also changes sign under the operation $\Phi^{\rm C}_{21} \rightarrow \Phi^{\rm C}_{21} + \pi$ and simultaneously, swapping the relationship between $\Phi^{\rm C}_{21}$ and $\Phi^{\rm C}_{22}$ from equation (\ref{eq:Phi_relation_oblate})  to (\ref{eq:Phi_relation_prolate}), or \emph{vice versa}, which simply amount to the transformation $\Phi^{\rm C}_{22} \rightarrow \Phi^{\rm C}_{22} + \pi$.    It follows that, in the absence of additional non-gravitational wave information,  we can carry out a gravitational wave search over the reduced ranges of {\bf either}
\begin{equation}
0 \le \psi_{\rm pol} < \pi/2 , \hspace{5mm} 0 \le \Phi^{\rm C}_{21} < 2\pi ,
\end{equation}
allowing for \emph{both} the oblate and prolate relations of equations (\ref{eq:Phi_relation_oblate}) (for oblate stars) and (\ref{eq:Phi_relation_prolate}) (for prolate stars)  in calculating 
$\Phi^{\rm C}_{22}(\Phi^{\rm C}_{21})$, {\bf or} 
\begin{equation}
0 \le \psi_{\rm pol} < \pi , \hspace{5mm} 0 \le \Phi^{\rm C}_{21} < 2\pi ,
\end{equation}
with only one or other (but not both) of the oblate and prolate relations of equations (\ref{eq:Phi_relation_oblate}) (for oblate stars) and (\ref{eq:Phi_relation_prolate}) in calculating $\Phi^{\rm C}_{22}(\Phi^{\rm C}_{21})$.  If either of these restricted parameter spaces is used in a search, and a detection is made, with parameters $(\psi_{\rm pol}, \Phi^{\rm C}_{21}, \Phi^{\rm C}_{22})$, three other equally acceptable solutions can be obtained through successive applications of the transformation
\begin{equation}
 \psi_{\rm pol} \rightarrow \psi_{\rm pol} + \pi/2, \hspace{5mm} \Phi^{\rm C}_{2m} \rightarrow \Phi^{\rm C}_{2m} + \pi .
 \end{equation}

Alternatively, in a search for a general triaxial body (as described in Section \ref{sect:tna_waveform}), finding a relationship between $\Phi^{\rm C}_{21}$ and $\Phi^{\rm C}_{22}$ of the form of either of equations  (\ref{eq:Phi_relation_oblate}) or (\ref{eq:Phi_relation_prolate})   would be a sign that the detected signal is coming from a biaxial star.  

See the second column of  Table \ref{table:waveform} for a summary of these possible choices in parameter ranges.

An alternative choice would have been to instead use equations (\ref{eq:C_22_positive}) and  (\ref{eq:C_21_positive}) for both the $\Delta I_{31} > 0$ and $\Delta I_{31} < 0$ cases, with the understanding that $C_{22}$  and $C_{21}$ can now be either positive or negative, but  both of the same sign (i.e.  both positive, or both negative).  Equations (\ref{eq:oblate_params_2})--(\ref{eq:Phi_relation_oblate}) then apply in both the oblate and prolate cases.  The waveform can the made made to change sign under the operation $C_{2m} \rightarrow -C_{2m}$ (applied simultaneously to both the $C_{22}$ and $C_{21}$).  It follows that, in the absence of other information, one can search over the reduced parameter ranges of {\bf either}
\begin{equation}
0 < \psi_{\rm pol} < \pi/2, \hspace{5mm}   - \frac{2\Omega^2}{r} \Delta I_{\rm max} < \tilde C_{2m} < \frac{2\Omega^2}{r} \Delta I_{\rm max},
\end{equation}
{\bf or}
\begin{equation}
0 < \psi_{\rm pol} < \pi, \hspace{5mm}   0 < \tilde C_{2m} < \frac{2\Omega^2}{r} \Delta I_{\rm max},
\end{equation}
If using such a reduced parameter ranges, other equally acceptable solutions can be generated through successive uses of the transformation
\begin{equation}
 \psi_{\rm pol} \rightarrow \psi_{\rm pol} + \pi/2, \hspace{5mm}  \tilde C_{2m} \rightarrow - \tilde C_{2m}  .
 \end{equation}
We mention this possibility as, while not fitting into the scheme of $C_{2m}$ being the modulus of a complex number, it has the advantage of possibly being easier to implement, as there is only one relation connecting the phases $\Phi^{\rm C}_{22}$ and $\Phi^{\rm C}_{21}$, regardless of whether the body is oblate or prolate, so some users may find it easier to integrate into their search method.

\subsection{Simplest case: A triaxial star, spinning about a principal axis} \label{sect:ta_waveform}

In this case we can set $\theta = 0$ in equations (\ref{eq:I_22_TNA}) and (\ref{eq:I_21_TNA}), so that $I^{21}=0$, while
\begin{equation}
I^{22} =      \sqrt{\frac{8\pi}{5}} \Delta I_{21} e^{-2i(\Omega t + \phi_0 + \psi)} ,
\end{equation}
so that
\begin{equation}
C^{\rm complex}_{22} = \sqrt{\frac{8\pi}{5}} \Delta I_{21} e^{-2i\psi   } .
\end{equation}
If we follow the convention used in Section \ref{sect:ta_source}  and choose to insist that $\Delta I_{21} >0$,  we can then immediately read-off
\begin{eqnarray}
\label{eq:simplest_C_22}
C_{22} &=& \sqrt{\frac{8\pi}{5}} \Delta I_{21} ,\\
\Phi_{22} &=& -2\psi   .
\end{eqnarray}
Using the relation of equation  (\ref{eqn:Phi_22_relation}) these equations can be inverted to give:
\begin{eqnarray}
\frac{\Omega^2 \Delta I_{21}}{r}  &=& \frac{1}{2} \tilde C_{22} ,\\
2(\phi_{{\rm gw},0}+\psi) &=& \Phi^C_{22}  . 
\end{eqnarray}
Note that the parameters $\phi_{{\rm gw}, 0}$ and $\psi$ are degenerate, as expected for this case.   In the event of a detection,   the measured values of $(\tilde C_{22}, \Phi^{\rm C}_{22})$ could be inserted into the above equations to compute the corresponding source parameters, which in this case are related in a very straight-forward way.   

We have already identified ranges in the source parameters that cover all possible physicality distinct stellar configurations; see equation (\ref{eq:ta_DI_and_phi_ranges}).  These immediately translate into the waveform parameter ranges:
\begin{equation}
0 < \tilde C_{22} <  \frac{2\Omega^2}{r} \Delta I_{\rm max} , \hspace{5mm} 0 \le \Phi^C_{22} < 2\pi . 
\end{equation}
These parameter ranges, together with those of equations (\ref{eq:full_Omega})--(\ref{eq:full_psi_pol}), will be wide enough to accommodate all possible non-gravitational wave priors.

However, from the point of view of carrying out a gravitational wave search without such prior information, these ranges are redundantly large.  The waveform changes sign under the operation $\psi_{\rm pol} \rightarrow \psi_{\rm pol}  + \pi/2$ and also under the operation $\Phi^C_{22}  \rightarrow \Phi^C_{22}  + \pi$, and so we can reduce the ranges in one or other (but not both) of those parameters:
\begin{eqnarray}
&& 0 \le \psi_{\rm pol} \le \pi/2, \hspace{5mm} 0 \le \Phi^C_{22} \le 2\pi, {\bf or}  \\
&& 0 \le \psi_{\rm pol} \le \pi, \hspace{8mm}  0 \le \Phi^C_{22} \le \pi.
\end{eqnarray}
The first choice is the one that reflects the choice traditionally made in gravitational wave searches (see e.g. \citet{known_pulsars_2014}).  In the event of a successful detection with parameters $(\psi_{\rm pol}, \Phi^C_{22})$, three other equally acceptable solutions can be generated by successive applications the of the transformation 
\begin{equation}
\psi_{\rm pol} \rightarrow \psi_{\rm pol}  + \pi/2, \hspace{5mm} \Phi^C_{22}  \rightarrow \Phi^C_{22}  + \pi .
\end{equation}

See the final column of Table \ref{table:waveform} for a summary of these possible choices in parameter ranges.

\section{Relation between the priors} \label{sect:priors}

As described above, in carrying out a gravitational wave search one has  a choice as to which set of variables are used, the source parameters or the waveform parameters.  If one is using Bayesian methods to conduct the search, one also needs to specify 
prior information on the range of each parameter, and supply a function giving one's initial belief as to its probability distribution.   For the triaxial star rotating about a principal axis, the two sets are essentially the same, but for the biaxial star, and the triaxial star not rotating about a principal axis, there is a non-trivial conversion to be made.  

Unfortunately, it is not obvious what a physically motivated choice of priors would be, in terms of either set of parameters.  This is  particularly true  for the amplitude-like source parameters $\Delta I_{21}, \Delta I_{31}$ or the wave parameters $\tilde C_{21}, \tilde C_{22}$.   The choice of priors would be related to the strength of the solid crust, the precise mechanism producing crustal deformation, and, for the  model of \citet{jone10}, the strength and orientation of the superfluid pinning.  We will therefore content ourselves here with a relatively simple consideration: if we make some simple choice of priors for the source parameters, we will evaluate the corresponding priors for the waveform parameters.  This will illustrate the fact that a simple choice of, say, a relatively simple rectangular-type distribution in one set of parameters, does not correspond to such a simple distribution when expressed in the other set.

We will only consider the biaxial case, as there it is easy to carry out calculations analytically.  The non-trivial conversion is between the `wobble angle' $\theta$ and the asymmetry $\Delta I_{31}$ for the source parameters, and the amplitudes $C_{21}$ and $C_{22}$ in for the waveform parameters.  The relevant formulae are:
\begin{eqnarray}
\label{eq:C_21_conversion}
C_{21} &=& \sqrt{\frac{8\pi}{5}} \Delta I_{31} \sin2\theta , \\
\label{eq:C_22_conversion}
C_{22} &=& \sqrt{\frac{8\pi}{5}} \Delta I_{31} \sin^2\theta .
\end{eqnarray}
We will take as a simple example of a set of priors the following: 
\begin{enumerate}
\item $\theta$ drawn by choosing a point randomly and uniformly  from the upper half of the unit sphere $0 \le \theta \le \pi/2$.  (The full range $0 \le \theta < \pi$ is not required, because of the degeneracy discussed in Section \ref{sect:biax_source}; we can always insist our $\theta$ value lies in this upper hemisphere).
\item $\Delta I_{31}$ drawn uniformly  over the interval $(0, \Delta I_{\rm max})$, independently of the value of $\theta$, corresponding to an oblate star.
\end{enumerate}
To simplify things, we can work with a dimensionless quantity $\hat I$:
\begin{equation}
\hat I \equiv \frac{\Delta I_{31}}{\Delta I_{\rm max}} .
\end{equation}
The corresponding separately normalised priors are
\begin{eqnarray}
P(\theta) &=& \sin\theta , \hspace{10mm} 0 \le \theta \le \pi/2, \\
P(\hat I) &=& 1, \hspace{15mm}  0 \le \hat I \le 1 ,
\end{eqnarray}
giving a joint prior
\begin{equation}
P(\theta, \hat I) = \sin\theta .
\end{equation}
The parameter space is simply a rectangle in $(\theta, \hat I)$ coordinates, with a probability density that depends only upon $\theta$.

To eliminate annoying factors, and made our amplitudes dimensionless, define
\begin{eqnarray}
\hat C_{21} &\equiv& \frac{C_{21}}{\sqrt{\frac{8\pi}{5}} \Delta I_{\rm max}} \\
\hat C_{22} &\equiv& \frac{C_{22}}{\sqrt{\frac{8\pi}{5}} \Delta I_{\rm max}} 
\end{eqnarray}
so that our transformation equations become
\begin{eqnarray}
\hat C_{21} &=& \hat I \sin2\theta , \\
\hat C_{22} &=&  \hat I \sin^2\theta .
\end{eqnarray}

To see the shape of the parameter space in the $(\hat C_{21}, \hat C_{22})$ variables we can look at the images of all four sides of the rectangle formed by the $(\theta, \hat I)$ variables:
\begin{enumerate}
\item The side $\theta=0$, $0 \le \hat I \le 1$ maps to $\hat C_{21} = \hat C_{22} = 0$, i.e.\ collapses to the origin.
\item The side $\hat I = 0$, $0 \le \theta \le \pi/2$ also collapses to the origin.
\item The side $\theta = \pi/2$, $0 \le \hat I \le 1$ maps to $\hat C_{21} = 0$, $\hat C_{22} = \hat I \Rightarrow 0 \le \hat C_{22} \le 1$.
\item The side $\hat I = 1$, $0 \le \theta \le \pi/2$ maps to $\hat C_{21} = \sin2\theta$, $\hat C_{22} = \sin^2\theta$.  This can be shown to be equivalent to the curve $\hat C_{21} = +2\sqrt{\hat C_{22}(1-\hat C_{22})}$. 
\end{enumerate}

To find the actual probability distribution within this closed region we can use the conversion formula
\begin{equation}
P(\hat C_{21}, \hat C_{22}) \det(J) = P(\theta)  P(\hat I) ,
\end{equation}
where $J$ is the Jacobian of the transformation:
\begin{equation}
J = \left( \begin{array}{cc} \frac{\partial \hat C_{21}}{\partial \hat I} &  \frac{\partial \hat C_{21}}{\partial \theta}  \\
\frac{\partial \hat C_{22}}{\partial \hat I} &  \frac{\partial \hat C_{22}}{\partial \theta}  \end{array} \right) 
=
\left( \begin{array}{cc} \sin2\theta &  \hat I 2\cos2\theta  \\
\sin^2\theta  &  \hat I \sin2\theta  \end{array} \right)  .
\end{equation}
Then
\begin{equation}
\det(J) = 2 \hat I \sin^2\theta = 2 \hat C_{22} .
\end{equation}
Eliminating $\Delta I_{31}$ between equations (\ref{eq:C_21_conversion}) and  (\ref{eq:C_22_conversion}) we have 
\begin{equation}
\sin\theta = \left[1+\left(\frac{\hat C_{21}}{2\hat C_{22}}\right)^2\right]^{-1/2} ,
\end{equation}
and so we obtain
\begin{equation}
\label{eq:pdf_waveform}
P(\hat C_{21}, \hat C_{22}) = \frac{1}{[\hat C_{21}^2+(2\hat C_{22})^2]^{1/2}} .
\end{equation}

\begin{figure} isa
\centerline{\includegraphics[height= 8cm,clip]{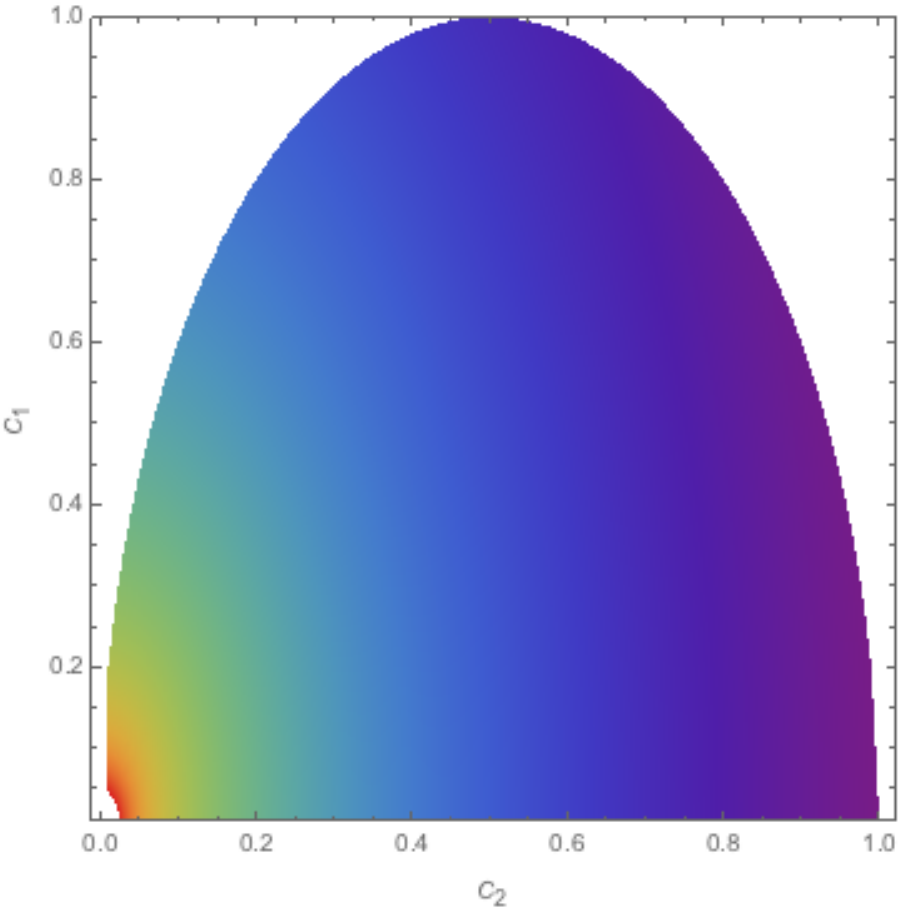}} \caption{Probability density function for the prior probabilities in  $(\hat C_{21}, \hat C_{22})$ coordinates corresponding to choice of priors in $(\theta, \Delta I_{31})$ described in the text, relevant to a biaxial star. We plot the logarithm of the probability distribution function of equation (\ref{eq:pdf_waveform}), and truncate the plot close to the (singular) origin.  \label{fig:C-params}}
\end{figure}

A plot showing the prior probability distribution as expressed in terms of  $(\hat C_{21}, \hat C_{22})$ is given in Figure \ref{fig:C-params}.   The relatively simple  probability distribution of the $(\theta, \hat I)$ coordinates, non-zero over a rectangular region,  has mapped into a highly non-uniform probability distribution, bounded by the  curves described above, with the probability density going singular  at the origin of the $(\hat C_{21}, \hat C_{22})$ system.  We plot the logarithm of the probability distribution to minimise contrast between different points, and truncate the plot close to the singular origin.  This serves to illustrate that a simple choice of priors in terms of one set of variables can lead to a more complex prior function in terms of the other set.

\section{Summary and discussion} \label{sect:summary}

We have examined various models for gravitational wave emission from steadily rotating neutron stars.  We have shown that, when written in term of the conventional parameters, termed here the `source parameters', the waveforms contain a number of \emph{discrete} degeneracies, with different values of the source parameters giving rise to exactly the same received gravitational wave signal $h(t)$.  These are related to arbitrary choices as to  how one lays down the axes $(O\tilde x, O\tilde y, O\tilde z)$ on the rotating star.  We show how these degeneracies can be removed by restricting the ranges of the source parameters, explicitly giving one (out of a possible multitude) of choices for each model.

We also re-examined the  gravitational wave emission model of \citet{jone10}, which described a star containing a pinned superfluid.  The waveform as written down in \citet{jone10} contained a \emph{continuous} degeneracy.  To remove this continuous degeneracy we introduced a new way of writing down the waveform in terms of so-called \emph{waveform parameters}.  Any future gravitational wave search for such a signal would be best performed using the waveform parameterisation.  However, in the event of a detection, it may be of interest to convert the waveform parameters back into a ($1$-parameter) family of source parameters.   The formulae given in this paper will allow such a conversion, and the minimal ranges in source parameters identified may help in avoiding unnecessary complications in this process.   We also briefly commented upon the (non-trivial) relation between prior probabilities assigned on one set of parameters verses priors on the other set.

The results of this paper will be of use to researchers devising methods to detect such mutli-harmoinc continuous gravitational wave signals, and in carrying out parameter estimation in the event of a detection.  A study of precisely these issues is underway, and will be presented elsewhere (Pitkin, Gill, Jones \& Woan, in preparation).

One other sort of continuos gravitational wave  search that has received attention concerns r-modes.  As shown in \citet{owen10}, the gravitational wave signal from a (current quadrupole-dominated) r-mode  is very similar to that of a (mass quadrupole-dominated) steadily spinning triaxial star, with a rotation of $\pi/4$ in $\psi_{\rm pol}$ transforming one class of signal into the other.  It follows that the discussion of triaxial aligned stars, of Section \ref{sect:ta_source}, applies to r-modes, with only minor modification; the quantity $\Delta I_{21}$ of the triaxial aligned star has only to be replaced by a measure the the current quadrupole of the r-mode.  The size of the current quadrupole is often measured by the dimensionless parameter $\alpha$; see e.g.\ equation (5) of \citet{owen10}.    If one mimics the convention used here, where we insist $\Delta I_{21} > 0$, to require $\alpha > 0$, the r-mode waveform has degeneracies only with respect to the polarisation angle $\psi_{\rm pol}$ and the overall phase of the signal $\Phi_{{\rm GW}, 0}$, and the simple degeneracies of section  \ref{sect:ta_source} apply again. 

We end with a few comments on electromagnetically-derived priors.  The spin frequency and sky location, derived from radio pulsar observations, are routinely used in \emph{targeted} gravitational wave searches, see e.g. \citet{aetal_11}.  For a small number of stars, information derived from pulsar wind nebulae has been used to provide values for the spin inclination angles $(\iota, \psi_{\rm pol})$.  The pulsar wind nebulae observations are in fact insensitive to the difference between the spin vectors $\bf \Omega$ and $- \bf \Omega$, which in the notation used here means they are insensitive to the transformation $(\iota, \psi_{\rm pol}) \rightarrow (\pi - \iota, \psi_{\rm pol}+ \pi)$, so the value of $\psi_{\rm pol}$ is determined only up to the addition of multiples of $\pi$.  This is sufficient to break the discrete $\psi_{\rm pol} \rightarrow \psi_{\rm pol} + \pi/2$ degeneracy considered in this paper, and so is sufficient break the discrete  degeneracy in the remaining source parameters.  This means that a successful gravitational wave detection that made use of the electromagnetic information on $\psi_{\rm pol}$  would provide unique values for all parameters  (aside from the continuous $1$-parameter degeneracy if the triaxial non-aligned model is being considered in terms of source parameters), apart from $\psi_{\rm pol}$, which would have the $\pi$ ambiguity.

It is difficult to see how prior information on other parameters  could be obtained, with the possible exception of the parameter $\phi_{{\rm gw}, 0}$.  This parameter essentially gives the rotational phase of the star's mass quadrupole at (retarded) time $t=0$.  In principle, one could try and use some theoretical model to relate the rotational phase of the star's mass quadrupole to some electomagnetically-observed feature.  There are (at least) two types of observation that one could potentially use in this regard.  Firstly, models where magnetic strains are responsible for deforming the star  and generating the non-zero time varying mass quadrupole (see e.g. \citet{mlm13})  might lead to a prediction for the relative phasing of the gravitational wave and pulsar signals, giving a prediction of the value of $\phi_{{\rm gw}, 0}$.  Secondly, \citet{bild98} proposed that temperature asymmetries in the star's crust could generate a non-zero mass quadrupole (see also \citet{UCB00}).  The observation of such a temperature asymmetry (as a modulation in surface brightness) could also be used to predict the absolute  the phasing of the gravitational wave signal, although (as far as we are aware) there are no candidate objects of this class of interest for gravitational wave searches.  

Note that in both cases it is necessary to ensure the model is sufficiently detailed to give the relative phasing of gravitational wave and electromagnetic signals, taking proper account of the fact that the gravitonal wave signal is sourced by an integral over the whole body's mass distribution (as per equation (\ref{eq:I_2m})), while the electromagnetic signals will be produced either at or above the surface.  In the case of making use of the pulsar radio phase, this is particularly problematic, as the radio pulsations are likely produced at some considerable altitude above the surface, and it is not clear if the large scale \emph{external} magnetic field, connected with the pulsations, will be aligned with the large scale \emph{internal} magnetic field, sourcing the mass quadrupole deformation.  

Rather than attempting to supply a prior on the rotational phase $\phi_{{\rm gw}, 0}$, a more pragmatic approach might be to be leave this phase unconstrained, and wait until a sufficiently large number of gravitational wave observations have been made.  A statistical analysis could then be used to see if this phase correlates with, say, the pulsar phase.  A correlation of this sort  would then be a sign (but not a proof)  that the magnetic fields do indeed play a role in significantly deforming the star, potentially providing a discriminant between magnetic-induced deformation and crustal strains, braking this degeneracy, and adding significantly to the physical insight provided by the gravitational wave observations.  Clearly, more detailed modelling is required to properly explore this interesting issue.

\section*{acknowledgments}

DIJ acknowledges support from STFC via grant number ST/H002359/1, travel support from NewCompStar (a COST-funded Research Networking Programme), and useful feedback from members of the Continuous Wave group of the LIGO Scientific Collaboration and the Virgo Scientific Collaboration, including many discussions with Matt Pitkin and Graham Woan.

\bibliography{references}

\begin{thebibliography}{}

\bibitem[\protect\citeauthoryear{{Aasi}, {Abadie}, {Abbott}, {Abbott},
  {Abbott}, {Abernathy}, {Accadia}, {Acernese}, {Adams}, {Adams} \& et
  al.}{{Aasi} et~al.}{2014}]{known_pulsars_2014}
{Aasi} J.,  {Abadie} J.,  {Abbott} B.~P.,  {Abbott} R.,  {Abbott} T.,
  {Abernathy} M.~R.,  {Accadia} T.,  {Acernese} F.,  {Adams} C.,  {Adams} T.,
   et al. 2014, \apj, 785, 119

\bibitem[\protect\citeauthoryear{{Aasi}, {Abbott}, {Abbott}, {Abbott},
  {Abernathy}, {Acernese}, {Ackley}, {Adams}, {Adams}, {Adams} \& et
  al.}{{Aasi} et~al.}{2015}]{LVC_narrowband}
{Aasi} J.,  {Abbott} B.~P.,  {Abbott} R.,  {Abbott} T.,  {Abernathy} M.~R.,
  {Acernese} F.,  {Ackley} K.,  {Adams} C.,  {Adams} T.,  {Adams} T.,    et al.
  2015, \prd, 91, 022004

\bibitem[\protect\citeauthoryear{{Abbott}, {Abbott}, {Adhikari}, {Ajith},
  {Allen}, {Allen}, {Amin}, {Anderson}, {Anderson}, {Arain} \& et al.}{{Abbott}
  et~al.}{2008}]{beating_crab_08}
{Abbott} B.,  {Abbott} R.,  {Adhikari} R.,  {Ajith} P.,  {Allen} B.,  {Allen}
  G.,  {Amin} R.,  {Anderson} S.~B.,  {Anderson} W.~G.,  {Arain} M.~A.,    et
  al. 2008, \apjl, 683, L45

\bibitem[\protect\citeauthoryear{{Andersson}, {Ferrari}, {Jones}, {Kokkotas},
  {Krishnan}, {Read}, {Rezzolla} \& {Zink}}{{Andersson}
  et~al.}{2011}]{aetal_11}
{Andersson} N.,  {Ferrari} V.,  {Jones} D.~I.,  {Kokkotas} K.~D.,  {Krishnan}
  B.,  {Read} J.~S.,  {Rezzolla} L.,    {Zink} B.,  2011, General Relativity
  and Gravitation, 43, 409

\bibitem[\protect\citeauthoryear{{Bejger} \& {Kr{\'o}lak}}{{Bejger} \&
  {Kr{\'o}lak}}{2014}]{bk14}
{Bejger} M.,  {Kr{\'o}lak} A.,  2014, Classical and Quantum Gravity, 31, 105011

\bibitem[\protect\citeauthoryear{{Bildsten}}{{Bildsten}}{1998}]{bild98}
{Bildsten} L.,  1998, \apjl, 501, L89

\bibitem[\protect\citeauthoryear{{Jaranowski}, {Kr{\'o}lak} \&
  {Schutz}}{{Jaranowski} et~al.}{1998}]{jks98}
{Jaranowski} P.,  {Kr{\'o}lak} A.,    {Schutz} B.~F.,  1998, \prd, 58, 063001

\bibitem[\protect\citeauthoryear{{Jones}}{{Jones}}{2010}]{jone10}
{Jones} D.~I.,  2010, \mnras, 402, 2503

\bibitem[\protect\citeauthoryear{{Jones}}{{Jones}}{2012}]{DCC_examples}
{Jones} D.~I.,  2012, Technical Report LIGO-T1200476, {Calculating
  gravitational waveforms: examples}

\bibitem[\protect\citeauthoryear{{Jones} \& {Andersson}}{{Jones} \&
  {Andersson}}{2001}]{ja01}
{Jones} D.~I.,  {Andersson} N.,  2001, \mnras, 324, 811

\bibitem[\protect\citeauthoryear{{Jones} \& {Andersson}}{{Jones} \&
  {Andersson}}{2002}]{ja02}
{Jones} D.~I.,  {Andersson} N.,  2002, \mnras, 331, 203

\bibitem[\protect\citeauthoryear{{Mastrano}, {Lasky} \& {Melatos}}{{Mastrano}
  et~al.}{2013}]{mlm13}
{Mastrano} A.,  {Lasky} P.~D.,    {Melatos} A.,  2013, \mnras, 434, 1658

\bibitem[\protect\citeauthoryear{{Owen}}{{Owen}}{2010}]{owen10}
{Owen} B.~J.,  2010, \prd, 82, 104002

\bibitem[\protect\citeauthoryear{{Thorne}}{{Thorne}}{1980}]{thor80}
{Thorne} K.~S.,  1980, Reviews of Modern Physics, 52, 299

\bibitem[\protect\citeauthoryear{{Ushomirsky}, {Cutler} \&
  {Bildsten}}{{Ushomirsky} et~al.}{2000}]{UCB00}
{Ushomirsky} G.,  {Cutler} C.,    {Bildsten} L.,  2000, \mnras, 319, 902

\bibitem[\protect\citeauthoryear{{Zimmermann} \& {Szedenits} Jr.}{{Zimmermann}
  \& {Szedenits}}{1979}]{zs79}
{Zimmermann} M.,  {Szedenits} Jr. E.,  1979, \prd, 20, 351

\end{thebibliography}

\appendix

\section{Rotations described in terms of Euler angles } \label{sect:rotations}

It will be useful to describe rotations about one of the body axes $(\tilde x, \tilde y, \tilde z)$ in terms of transformations of the Euler angles $(\theta, \phi, \psi)$.  To do so, it is useful to write down the unit vectors along the body axes $({\bf e_{\tilde x}}, {\bf e_{\tilde y}}, {\bf e_{\tilde z}})$  in terms of their components with respect to the inertial frame unit vectors.  This is most easily done by constructing the matrix that carries out an active rotation from inertial axes to body axes.   We need to first perform a rotation of $\psi$ about the inertial $z$-axis, then a rotation of $\theta$ about the inertial $x$-axis, and finally a rotation of $\phi$ about the inertial $z$-axis.  In an obvious notation, we then have
\begin{equation}
R_{ab} = [R^z(\phi) R^x(\theta) R^z(\psi)]_{ab} ,
\end{equation}
The relevant active rotation matrices are:
\begin{eqnarray}
R_{ab}^z(\psi) &=& 
\left[ \begin{array}{ccc} \cos\psi & -\sin\psi& 0 \\  \sin\psi & \cos\psi & 0 \\ 0 & 0 & 1 \end{array} \right] \\
R_{ab}^x(\theta) &=& 
\left[ \begin{array}{ccc} 1 & 0 & 0 \\  0 & \cos\theta & -\sin\theta \\ 0 & \sin\theta & \cos\theta \end{array} \right] \\
R_{ab}^z(\phi) &=& 
\left[ \begin{array}{ccc} \cos\phi & -\sin\phi& 0 \\  \sin\phi & \cos\phi & 0 \\ 0 & 0 & 1 \end{array} \right] 
\end{eqnarray}
Allowing $R_{ab}$ to act on the inertial unit vectors $({\bf e_{x}}, {\bf e_{y}}, {\bf e_{z}})$, we obtain
\begin{eqnarray}
{\bf e_{\tilde x}} &=& \left[ \begin{array}{c}
\cos\phi \cos\psi  -  \sin\phi \cos\theta \sin\psi \\ 
\sin\phi \cos\psi  +  \cos\phi \cos\theta \sin\psi \\ 
\sin\theta \sin\psi 
\end{array} \right] \\
{\bf e_{\tilde y}} &=& \left[ \begin{array}{c}
-\cos\phi \sin\psi  -  \sin\phi \cos\theta \cos\psi \\ 
-\sin\phi \sin\psi  +  \cos\phi \cos\theta \cos\psi \\ 
\sin\theta \cos\psi 
\end{array} \right] \\
{\bf e_{\tilde z}} &=& \left[ \begin{array}{c}
\sin\phi \sin\theta \\ 
-\cos\phi \sin\theta \\ 
\cos\theta 
\end{array} \right] 
\end{eqnarray} 
We can then see how these body-frame basis vectors transform under operations of the form $(\theta, \phi, \psi) \rightarrow (\hat\theta, \hat\phi, \hat\psi)$ in order to identify the nature of the corresponding rotation of the body axes.

\subsection{Rotation of $\pi$ about $O\tilde z$.} \label{sect:pi_about_z}

Under $\psi \rightarrow \psi + \pi$, we see that
\begin{eqnarray}
{\bf e_{\tilde x}} &\rightarrow& - {\bf e_{\tilde x}} , \\
{\bf e_{\tilde y}} &\rightarrow& - {\bf e_{\tilde y}} , \\
{\bf e_{\tilde z}} &\rightarrow&  + {\bf e_{\tilde z}} .
\end{eqnarray}
This is clearly a rotation of $\pi$ about $O\tilde z$.  

\subsection{Rotation of $\pi$ about $O\tilde y$.} \label{sect:pi_about_y}

Now consider the transformation
\begin{eqnarray}
\theta &\rightarrow& \pi-\theta , \\
\phi &\rightarrow& \phi + \pi , \\
\psi &\rightarrow& -\psi . 
\end{eqnarray}
This produces the transformation
\begin{eqnarray}
{\bf e_{\tilde x}} &\rightarrow& - {\bf e_{\tilde x}} , \\
{\bf e_{\tilde y}} &\rightarrow& + {\bf e_{\tilde y}} , \\
{\bf e_{\tilde z}} &\rightarrow&  - {\bf e_{\tilde z}} .
\end{eqnarray}
This is clearly a rotation of $\pi$ about $O\tilde y$.

\subsection{Rotation of $\pi/2$ about $O\tilde y$.} \label{sect:pi_by_2_about_y}

For a rotation of $\pi/2$ about $O\tilde y$, we require
\begin{eqnarray}
{\bf e_{\tilde x}} &\rightarrow& - {\bf e_{\tilde z}} , \\
{\bf e_{\tilde y}} &\rightarrow& + {\bf e_{\tilde y}} , \\
{\bf e_{\tilde z}} &\rightarrow&  + {\bf e_{\tilde x}} .
\end{eqnarray}
To see what transformation in the Euler angles produces this, we need to solve the equations
\begin{eqnarray}
{\bf e_{\tilde x}}(\hat\theta, \hat\phi, \hat\psi) &=& - {\bf e_{\tilde z}} (\theta, \phi, \psi)  , \\
{\bf e_{\tilde y}}(\hat\theta, \hat\phi, \hat\psi)  &=& + {\bf e_{\tilde y}} (\theta, \phi, \psi)  , \\
{\bf e_{\tilde z}}(\hat\theta, \hat\phi, \hat\psi)  &=&  + {\bf e_{\tilde x}} (\theta, \phi, \psi)  .
\end{eqnarray}
The solution does not correspond to a simple translation or sign change for the Euler angles, so we give the results for the various sine and cosine functions instead:
\begin{eqnarray}
\label{eq:sin_theta_tilde}
\sin\tilde\theta &=&  (1-\sin^2\theta \sin^2\psi)^{1/2} , \\
\cos\tilde\theta &=& \sin\theta \sin\psi \\
\sin\tilde\phi &=& - \frac{-\cos\phi \cos\psi + \sin\phi \cos\theta \sin\psi}{(1-\sin^2\theta \sin^2\psi)^{1/2}} , \\
\cos\tilde\phi &=& - \frac{\sin\phi \cos\psi + \cos\phi \cos\theta \sin\psi}{(1-\sin^2\theta \sin^2\psi)^{1/2}} , \\ 
\sin\tilde\psi &=& - \frac{\cos\theta}{(1-\sin^2\theta \sin^2\psi)^{1/2}} , \\
\label{eq:cos_psi_tilde}
\cos\tilde\psi &=&  \frac{\sin\theta \cos\psi}{(1-\sin^2\theta \sin^2\psi)^{1/2}} .
\end{eqnarray}

\section{The transform \lowercase{$h^{\Omega, 2\Omega}_{+, \times} \rightarrow - h^{\Omega, 2\Omega}_{+, \times}$ }} \label{sect:h_to_minus_h}

As described in Section \ref{sect:source_params}, it is useful to identify parameter transformations such that $h^{\Omega, 2\Omega}_{+, \times} \rightarrow - h^{\Omega, 2\Omega}_{+, \times}$, as these combined with transformations of the form $\psi_{\rm pol} \rightarrow \psi_{\rm pol} + \pi/2$ give rise to a to give a symmetry in the waveform, allowing for a reduction in the parameter space range.  

As described in Section \ref{sect:waveform_params}, in terms of  the waveform parameters, the transformation is simply
\begin{equation}
\Phi^{\rm C}_{2m} \rightarrow \Phi^{\rm C}_{2m} + \pi .
\end{equation}

In terms of the source parameters,  in the case of a triaxial aligned star, the transformation can be achieved straightforwardly, by the mapping $(\phi_{{\rm gw},0} + \psi)  \rightarrow (\phi_{{\rm gw},0} + \psi) + \pi/2$, as described in equation (\ref{eq:transform_ta_source}).  In the case of a biaxial star, the transformation is again straightforward, as given by the mapping $\Delta I_{31} \rightarrow -\Delta I_{31}$ of equation (\ref{eq:transform_biaxial_source}), corresponding to a transformation of the form  $\delta \rho \rightarrow - \delta \rho$. 

However, in the case of a triaxial non-aligned star, our choice of following the convention common in ridge body dynamics of setting  $I_3 > I_2 > I_1$ makes the transformation $\delta \rho \rightarrow - \delta \rho$ more difficult to describe.    Suppose the  original star density perturbation $\delta \rho$  produces perturbations in the moment of inertia tensor $\delta I_1, \delta I_2, \delta I_3$, so that  the moment of inertia tensor, referred to the body axes, is
\begin{eqnarray}
I_1 &=& I_0 + \delta I_1 , \\
I_2 &=& I_0 + \delta I_2 , \\
I_3 &=& I_0 + \delta I_3 , 
\end{eqnarray}
where we choose the axes such that $I_3 > I_2 > I_1$.  We are always free to make this choice, and it means that our asymmetry parameters then satisfy  $\Delta I_{31} > \Delta I_{21} > 0$, such that the largest principal part of the moment of inertia tensor lies along $O\tilde z$, and the smallest along $O\tilde x$.   Given this convention,  we can't simply flip the signs of $\Delta I_{31}$ and $\Delta I_{21}$.  We need to find a more complicated transformation, consistent with out convention $I_3 > I_2 > I_1$. 

Under the operation $\delta \rho \rightarrow -\delta  \rho$, the moment of inertia tensor transforms to  
\begin{eqnarray}
I_1 &=& I_0 - \delta I_1 , \\
I_2 &=& I_0 - \delta I_2 , \\
I_3 &=& I_0 - \delta I_3 .
\end{eqnarray}
The $O\tilde x$ axis is now the axis of largest principal moment of inertia, and the $O\tilde z$ the smallest.  To put this star into our conventional form we need to rotate by $\pm \pi/2$ about the $O\tilde y$ axis.  We also need to choose the new perturbations in the moment of inertia tensor to be
\begin{eqnarray}
\delta \hat I_1 &=& - \delta I_3 , \\
\delta \hat I_2 &=& - \delta I_2 , \\
\delta \hat I_3 &=& - \delta I_1 , 
\end{eqnarray}
so that the new amplitude parameters are now related to the old by
\begin{eqnarray}
\Delta \hat I_{21} &\equiv& \delta \hat I_2 - \delta \hat I_1 = -\delta  I_2 + \delta  I_3 
= \Delta I_{31} - \Delta I_{21}  , \\
\Delta \hat I_{31} &\equiv& \delta \hat I_3 - \delta \hat I_1 = -\delta  I_1 + \delta  I_3 
= \Delta I_{31} . 
\end{eqnarray}
Note that with this choice we have $\Delta \hat I_{31} > \Delta \hat I_{21} > 0$, as required by our convention.  If we then carry out the transformation
\begin{eqnarray}
\label{eq:euler_transform}I_1
(\theta, \phi, \psi)  &\rightarrow& (\hat\theta, \hat\phi, \hat\psi) , \\
\label{eq:Delta_I_21_transform}
\Delta I_{21} &\rightarrow& \hat\Delta I_{21} = \Delta I_{31} - \Delta I_{21} , \\
\label{eq:Delta_I_31_transform}
\Delta I_{31} &\rightarrow& \hat\Delta I_{31} =  \Delta I_{31} 
\end{eqnarray}
where the new Euler angles $(\hat\theta, \hat\phi, \hat\psi)$ are those of Section \ref{sect:pi_by_2_about_y}, equations (\ref{eq:sin_theta_tilde})--(\ref{eq:cos_psi_tilde}), appropriate to a $\pi/2$ rotation about $O\tilde y$, we find $h^{\Omega, 2\Omega}_{+, \times} \rightarrow - h^{\Omega, 2\Omega}_{+, \times}$, as required.

\section{Summary tables of parameter ranges}

\begin{table*}
\begin{minipage}{115mm}
\caption{Table giving one possible choice for the ranges in the source parameters $(\theta, \phi_{{\rm gw},0}, \psi, \Delta I_{21}, \Delta I_{31})$ that uniquely label the orientational of \emph{any} star's mass quadrupole.
The  ranges in spin frequency, sky location and inclination angle, common to all models, are given in equations  (\ref{eq:full_Omega})--(\ref{eq:full_psi_pol}).  Note that in the triaxial aligned case, the parameters $\phi_{{\rm gw},0}$ and $\psi$ are degenerate, appearing only as the sum $(\phi_{{\rm gw},0} + \psi)$.   In the case where electromagnetic observations have \emph{not} provided a value for $\psi_{\rm pol}$, the range in this angle can be restricted from the $(0,\pi)$ interval of equation (\ref{eq:psi_pol_pi}) to the interval $(0,\pi/2)$.  If this is done, then, in the event of a gravitational wave detection, three other equally acceptable solutions can be obtained, each differing by the transformations given in equations (\ref{eq:transform_tna1_source}) and (\ref{eq:euler_transform})--(\ref{eq:Delta_I_31_transform})  for the triaxial non-aligned case, equation  (\ref{eq:transform_biaxial_source}) for the biaxial case,  and equation (\ref{eq:transform_ta_source}) for the triaxial aligned case.   \label{table:source}}
\begin{tabular}{c|c|c|c}
& Triaxial non-aligned & Biaxial & Triaxial aligned   \\
\hline
$\theta$ & $(0,\pi/2)$ & $(0,\pi/2)$&  $\theta=0$   \\
$\phi_{{\rm gw},0}$ &  $(0,2\pi)$ & $(0,2\pi)$  & $(0,\pi)$  \\
$\psi$ & $(0,\pi)$  & Not present & Degenerate with $\phi_{{\rm gw},0}$   \\
$\Delta I_{21}$ & $(0,\Delta I_{31})$  & $\Delta I_{21} = 0$ &$(0,\Delta I_{\rm max})$    \\
$\Delta I_{31}$ &  $(\Delta I_{21},\Delta I_{\rm max})$   & $(-\Delta I_{\rm max}, \Delta I_{\rm max}) $ &  Not present 
\end{tabular}
\end{minipage}
\end{table*}

\begin{table*}
\begin{minipage}{115mm}
\caption{Table giving one possible choice for the ranges in the waveform  parameters $\{C_{22}, C_{2}, \Phi^{\rm C}_{22}, \Phi^{\rm C}_{21} \}$, consistent with \emph{any} physical stellar configuration.  The  ranges in spin frequency, sky location and inclination angle, common to all models, are given in equations  (\ref{eq:full_Omega})--(\ref{eq:full_psi_pol}).   In the case where electromagnetic observations have \emph{not} provided a value for $\psi_{\rm pol}$, the range in this angle can be restricted from the $(0,\pi)$ interval of equation (\ref{eq:psi_pol_pi})  to the interval $(0,\pi/2)$.  In the event of a gravitational wave detection, three other equally acceptable solutions can be obtained, each differing by the transformations $\psi_{\rm pol} \rightarrow \psi_{\rm pol} = \pi/2$, $\Phi_{2m}^{\rm C} \rightarrow \Phi_{2m}^{\rm C}  + \pi$, as described in Section  \ref{sect:waveform_params}.   \label{table:waveform}}
\begin{tabular}{c|c|c|c}
&  Triaxial non-aligned & Biaxial & Triaxial aligned  \\
\hline
$C_{22} \sqrt{\frac{5}{8\pi}}$ &  $(0,\Delta I_{\rm max})$ & $(0,\Delta I_{\rm max})$ &  $(0,\Delta I_{\rm max})$   \\
$C_{21}  \sqrt{\frac{5}{8\pi}}$ & $(0,\Delta I_{\rm max})$    & $(0,\Delta I_{\rm max})$ & $C_{21}=0$  \\ 
$\Phi^C_{22}$ & $(0,2\pi)$  & $\Phi^C_{22} = 2 \Phi^C_{21} + \pi$ (oblate stars) & $(0,2\pi)$  \\
 &   & $\Phi^C_{22} = 2 \Phi^C_{21}$ (prolate stars) &   \\
$\Phi^C_{21}$ &  $(0,2\pi)$     & $(0,2\pi)$ &  Not present
\end{tabular}
\end{minipage}
\end{table*}

\end{document}